\documentclass[12pt,preprint]{aastex}

\shorttitle{Deuteration pathways in disks}
\shortauthors{Aikawa et al.}

\begin{document}


\title{MULTIPLE PATHS OF DEUTERIUM FRACTIONATION IN PROTOPLANETARY DISKS}


\author{Yuri Aikawa}
\affil{Department of Astronomy, The University of Tokyo}
\email{aikawa@astron.s.u-tokyo.ac.jp}
\author{Kenji Furuya}
\affil{Center for Computational Sciences, University of Tsukuba, Japan}
\author{Ugo Hincelin}
\affil{Department of Chemistry, The University of Virginia, USA}
\and
\author{Eric Herbst}
\affil{Department of Chemistry, The University of Virginia, USA}



\begin{abstract}

We investigate deuterium chemistry coupled with the nuclear spin-state chemistry of H$_2$ and H$_3^+$ in protoplanetary disks. 
Multiple paths of deuterium fractionation are found; exchange reactions with D atoms, such as HCO$^+$ + D, are effective
in addition to those with HD.
In a disk model with grain sizes appropriate for dark clouds, the freeze-out of molecules is severe in the outer midplane, 
while the disk surface is shielded from UV radiation. Gaseous molecules, including DCO$^+$, thus become abundant at the disk 
surface, which tends to make their column density distribution relatively flat.
If the dust grains have grown to millimeter size, the freeze-out rate of neutral species is reduced, 
and the abundances of gaseous molecules, including DCO$^+$ and N$_2$D$^+$, are enhanced in the cold midplane. 
Turbulent diffusion transports D atoms and radicals at the disk surface to the midplane, and stable ice species in the
midplane to the disk surface. The effects of turbulence on chemistry are thus multifold; while DCO$^+$ and N$_2$D$^+$ abundances 
increase or decrease depending on the regions, HCN and DCN in the gas and ice are much reduced at the innermost radii, compared 
with the model without turbulence. When cosmic rays penetrate the disk, 
the ortho-to-para ratio (OPR) of H$_2$ is found to be thermal in the disk, except in the cold ($\lesssim 10$ K) midplane. 
We also analyze the OPR of H$_3^+$ and H$_2$D$^+$, as well as the main reactions of H$_2$D$^+$, DCO$^+$, and N$_2$D$^+$ to analytically
derive their abundances in the cold midplane.

\end{abstract}


\keywords{astrochemistry --- star-formation --- protoplanetary disks}



\section{INTRODUCTION}

In the primordial material in the Solar System, such as comets and meteorites, molecular D/H ratios are often higher than
the elemental D/H ratio, which is $2\times 10^{-5}$ \citep{geiss98}. For example, the D/H ratio of water is found to be
$(1.6-5)\times 10^{-4}$ in comets \citep[e.g.][]{mumma11,altwegg15}. Since the zero-point energy of D-bearing species is
lower than that of the normal isotope by up to a few hundred K, the significant deuterium
enrichment is considered to originate in chemistry at low temperatures. 

There are two possible sites of such low-temperature chemistry: molecular clouds prior to star formation and outer regions of protoplanetary disks.
In molecular clouds, the temperature is $\sim 10$ K and high D/H ratios are found for various molecules. For example, 
even a doubly deuterated molecule, D$_2$CO, has been detected in prestellar cores, and the $n$(D$_2$CO)/$n$(H$_2$CO) ratio has been derived to be $0.01-0.1$
\citep{bacmann03}, where $n$($i$) denotes the number density of
species $i$. In such dense cold cores, where HD is the primary reservoir of deuterium,
deuterated H$_3^+$ is produced via exothermic exchange reactions; e.g.,
\begin{equation}
  {\rm H_3^+ + HD \rightarrow H_2D^+ + H_2}. \label{h2dp}
\end{equation}
The high D/H ratio
of H$_3^+$ propagates to other molecules via ion-molecule reactions in the gas phase. 
The atomic D/H ratio is also enhanced through the electron recombination of H$_3^+$ and H$_2$D$^+$. The high atomic D/H ratio propagates to icy molecules
through hydrogenation/deuteration of atoms and molecules on grain surfaces. 
The ions H$_2$D$^+$ and D$_2$H$^+$ are actually detected at the center of the prestellar core L1544
\citep{caselli03,vastel06}. In such a dense cold region, the D/H ratio of H$_3^+$ is further enhanced by the freeze-out of CO, which is otherwise the major reactant of H$_3^+$ and its isotopologs.

Protoplanetary disks are also partially ionized by X-rays and/or cosmic rays, and the temperature is $\lesssim 20$ K in the outer midplane region, so that
deuterium enrichment can proceed in disks, as well.
In fact, DCN, DCO$^+$, and N$_2$D$^+$ have been  detected in several disks \citep[e.g.][]{vandishoeck03, qi08, huang15}. While a stable neutral
species such as DCN can originate in interstellar ice, then delivered to and desorbed in the disk, molecular ions are apparently
formed in situ, considering their short destruction timescales. \cite{vandishoeck03} used the JCMT telescope to derive a disk-averaged
$n$(DCO$^+$)/$n$(HCO$^+$) ratio of 0.035 in TW Hya, providing evidence of ongoing deuterium enrichment in protoplanetary disks.
Although the deuterium enrichment in the primordial material in the Solar System can originate in interstellar chemistry,
at least partially, it has been debated if and to what extent the molecular D/H ratios are modified in disks \citep{aikawa99,willacy07,cleeves14}.
Spatially resolved observations of deuterated ions can specify the region of active deuteration in protoplanetary disks.

The emission of DCO$^+$ is  spatially resolved in the disks around TW Hya \citep{qi08} and HD 163296 \citep{mathews13}.
In both disks, the DCO$^+$ emission shows a ring structure, which is considered to reflect the production of DCO$^+$ by the reaction
\begin{equation}
{\rm H_2D^+ + CO \rightarrow DCO^+ + H_2}. \label{dcop}
\end{equation}
The exchange reaction (\ref{h2dp}) to form H$_2$D$^+$ is exothermic only by $\lesssim 250$ K, and thus its backward reaction becomes active at $T\gtrsim 30$ K.
\cite{mathews13} found that DCO$^+$ emission shows a ring structure of radius $r= 110-160$ au, and argued that the outer radius of the DCO$^+$
ring corresponds to the CO snow line. The ion DCO$^+$ is expected to be abundant in the region with temperature $T\sim 19-21$ K;
at higher temperatures ($T\gtrsim 21$ K), CO and ortho-H$_2$ (see below) would become abundant enough to destroy the precursor molecules, H$_3^+$ and H$_2$D$^+$, while
at $T\lesssim 19$ K, CO abundance would be too low to make DCO$^+$ abundant \citep{mathews13}.
The DCN emission in TW Hya disk, on the other hand, is centrally peaked.
This molecule is expected to form primarily via the reaction of N atoms with CHD, which is formed by the dissociative recombination of  CH$_2$D$^+$. Since the exothermicity of the exchange reaction
\begin{equation}
{\rm CH_3^+ + HD \rightarrow CH_2D^+ + H_2} \label{ch2dp}
\end{equation}
is higher than that of reaction (\ref{h2dp}), DCN can be abundant even at $T\gtrsim 30$ K \citep{MBH89,oberg12}.

Recent observations of deuterated species in disks, however, challenge the above scenario. \cite{qi15} observed DCO$^+$ in HD 163296 with a higher spatial resolution,
and found that the inner edge of the DCO$^+$ ring is at 40 AU, which is much closer to the central star than derived by \cite{mathews13}. 
\cite{huang17} observed six protoplanetary disks, spatially resolving DCO$^+$, H$^{13}$CO$^+$, H$^{13}$CN, and DCN in most of them.
While the DCO$^+$ emission tends to be spatially more extended than the DCN emission, the relative distributions of
DCO$^+$ and DCN vary among the disks. While the DCN emission is more compact than the DCO$^+$ emission in HD 163296, 
the radial intensity profiles of DCO$^+$ and DCN are similar in AS 209, for example. Such variations indicate that multiple paths
to deuteration occur in the disks.

On the theoretical side, deuterium chemistry in protoplanetary disks has long been investigated \citep[e.g.][]{aikawa99, willacy07, cleeves14}.
Recent progress in the study of deuterium fractionation includes the evaluation of the state-to-state rate coefficients of the H$_3^+$ + H$_2$ system \citep{hugo09}
and the re-evaluation of the exothermicity of reaction (\ref{ch2dp}) \citep{roueff13}. Since the ground-state
energy of ortho-H$_2$ (o$\mathchar`-$H$_2$) is higher than that of para-H$_2$ (p$\mathchar`-$H$_2$), the backward reactions of exchange reactions such as (\ref{h2dp}) are less
endothermic with o$\mathchar`-$H$_2$ than with p$\mathchar`-$H$_2$. The deuterium enrichment would thus be less efficient  if o$\mathchar`-$H$_2$ is abundant.
Although several groups have presented the deuterium chemistry in disks coupled with ortho-para chemistry,
they concentrated mostly on  molecular species such as HDO rather than DCO$^+$ and DCN, or made an assumption that the ortho-para chemistry
is locally thermalized \citep{albertsson14, cleeves14, cleeves16}. 
One of the exceptions is the paper by \cite{teague15}; these authors observed HCO$^+$ and DCO$^+$ lines in DM Tau, and calculated molecular evolution in a disk model to be compared with their observations.
While they solved for the spin states of H$_2$ and H$_3^+$, they did not seem to adopt the updated exothermicity for reaction (\ref{ch2dp}).
\cite{roueff13} calculated the exothermicity of reaction (\ref{ch2dp}) for p-H$_2$ to be 654 K, which is higher than
the previous estimate of 370 K \citep{smith82}.
\cite{favre15} adopted this updated energy difference for reaction (\ref{ch2dp}) in their chemical model, and showed that it enhances
the abundance of DCO$^+$ in the warm surface layers of the disk. Their model, however, does not include the ortho-para chemistry,
and implicitly assumes that all H$_2$ is in the para state. The assumption would not be appropriate in the warm layers.

In the present work,  we investigate the deuteration paths of HCO$^+$, N$_2$H$^+$, and HCN in protoplanetary disks using an updated
chemical reaction network with deuterium fractionation and ortho-para chemistry. Instead of constructing  best-fit models of
observed disks, we investigate the chemistry in template disk models. The effects of elemental C/O ratio, grain size, vertical mixing, and exclusion of cosmic rays are also studied. 
The rest of the paper is organized as follows. 
Our chemical reaction network and protoplanetary disk models are described in \S 2. Section 3 presents the results of our numerical calculations; i.e., 
abundance and column density distributions of HCO$^+$, N$_2$H$^+$, HCN and their deuterated isotopologs in disk models.
The model results are qualitatively compared with observations in \S 4, and our conclusions are summarized in \S 5.
In appendices, we also present an analysis of the OPRs of H$_2$, H$_3^+$, and H$_2$D$^+$, and  analytical formulae of abundances of
H$_2$D$^+$, DCO$^+$ and N$_2$D$^+$ in the cold midplane.

\section{MODELS}

We adopt basically the same disk structure and chemical reaction network as in \cite{aikawa15}. Although basic descriptions of the model and network
are given in this section, more detailed descriptions can be found in \cite{aikawa15}, \cite{furuya13}, and \cite{aikawa06}.

\subsection{Disk Models}


In the present work, we investigate deuterium chemistry in template disk models, rather than constructing a model for a specific object.
The radial distribution of the gas column density in our models is thus rather arbitrary; it is determined so that the mass accretion rate 
($10^{-8} M_{\odot}$/yr) is constant at all radii for a constant viscosity parameter $\alpha$ (see \S 3.3). Note that we do not explicitly take into account the
radial accretion of gas and dust in solving the chemical reaction network.
The central star is assumed to be a T Tauri star of mass $M_*$=0.5 $M_{\odot}$, surface temperature $T_*= 4000$ K, and radius $R_*=0.5 R_{\odot}$.
We refer to the UV and X-ray spectrum of TW Hya, and adopt the luminosities of $L_{\rm UV}=10^{31}$ erg s$^{-1}$ and
$L_{\rm X}=10^{30}$ erg s$^{-1}$ \citep{herczeg02, kastner02}\citep[see also][]{nomura05, nomura07}.
The vertical density distribution is set by hydrostatic equilibrium. The temperatures of gas and dust are obtained by solving two-dimensional 
radiative transfer and the balance between cooling and heating. The temperature and density structures are solved self-consistently.
The gas and dust temperatures are the same in the midplane, while gas is
warmer than dust at the disk surface, which is basically a dense photon-dominated region.

\begin{figure}
\includegraphics[scale=0.7, angle=0]{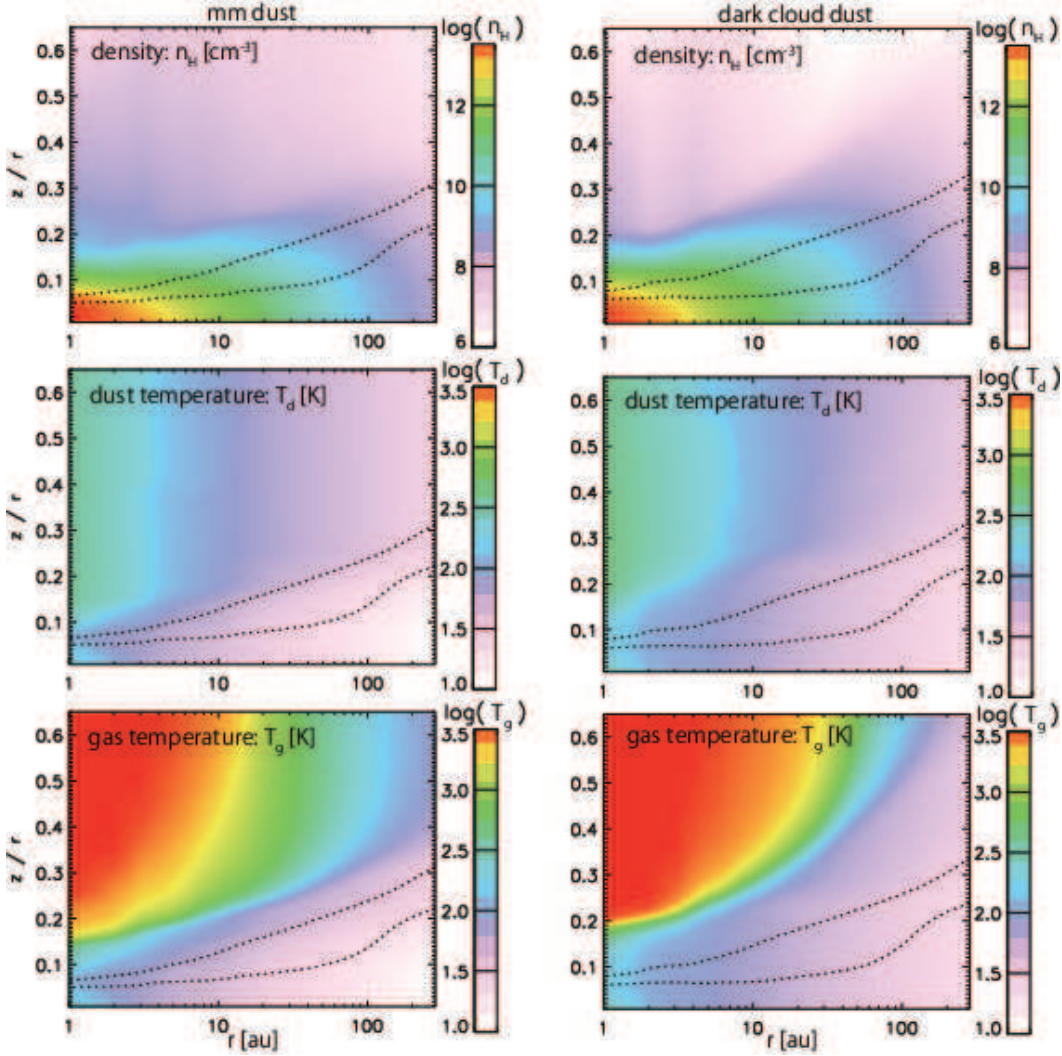}
\caption{Distributions of gas density (top), dust temperature (middle), and gas temperature (bottom) in the disk model with mm-sized grains (left) and dark cloud dust (right).
The upper dotted line depicts the positions where the X-ray ionization rate is equal to the cosmic-ray ionization rate ($5\times 10^{-17}$ s$^{-1}$), while the X-ray ionization rate
is $1\times 10^{-18}$ s$^{-1}$ at the lower dotted line.
\label{dist_phys}}
\end{figure}

In our fiducial model, the grains have grown to the maximum size $a_{\rm max}$ of 1 mm size, while the minimum size is $a_{\rm min}=0.01$  $\mu$m, with the dust-to-gass mass ratio of 0.01, all over the disk.
A grain size distribution is assumed to follow the power law $dn(a)/da\propto a^{-3.5}$, where $n(a)$ is the number density of grains with size $a$.
For comparison, we also adopt a disk model with dust appropriate to dark clouds, which we label dark cloud dust.  The maximum grain size of the latter model is 10 $\mu$m 
\citep{weingartner01}.
The density and temperature distributions of the two disk models are shown in Figure \ref{dist_phys}. The temperature is generally higher in the model with dark cloud dust,
because it  absorbs stellar radiation more efficiently. While we assume that the disk is static (i.e. no diffusion or accretion), 
we also calculate models with vertical mixing \citep{furuya13}, in order to investigate its effect on deuterium chemistry, 

\subsection{Chemical Model}

The gas-grain chemical reaction network is based on \cite{garrod06}, but has been updated to include reactions that are effective at high temperatures ($T\gtrsim 100$ K)
\citep{harada10,harada12}. Our network includes up to triply deuterated species, and nuclear spin states of H$_2$, H$_3^+$, and their isotopologs \citep{hugo09,hincelin14, coutens14, furuya15}.
We refer to \cite{roueff13} for the reaction rate coefficients and exothermicities/endothermcities of the exchange reaction (\ref{ch2dp}) 
and analogous reactions with multi-deuteration and spin states. 

Ultra violet radiation from the central star and interstellar radiation field causes photodissociation and photoionization in the gas phase; the rate coefficients are calculated by convolving
the dissociation/ionization cross sections of molecules and the UV spectrum at each position in the disk.
Self- and mutual shielding of H$_2$, HD, CO, N$_2$, and C atoms are taken into account \citep{draine96, visser09, wolcott11, li13,  kamp00}. 
The shielding of D$_2$ is not considered in the present model,
but we confirmed that it does not affect our results, because D$_2$ is much less abundant than HD at the disk surface, where photodissociation is effective.
Photo-dissociation of icy molecules is also considered in our model \citep[e.g.][]{furuya13}.
The ionization sources in the disk are X-rays, cosmic rays, and the decay of radioactive nuclei. We assume a cosmic-ray ionization rate of $5\times 10^{-17}$ s$^{-1}$
with an attenuation length of 96 g \citep{umebayashi81}. Considering that the penetration of cosmic rays can be prohibited by stellar winds, we also run a model without
cosmic rays  \citep{aikawa99b,cleeves14b}; the disk midplane is then mainly ionized by the decay of radioactive nuclei with a rate of $10^{-18}$ s$^{-1}$.
Layers above the midplane are mainly ionized by stellar X-rays; the ionization rate could reach $10^{-14}$ s$^{-1}$ at the disk surface, for example
\citep[see Figure 1 in][]{aikawa15}.

We adopt a two-phase model, which consists of a gas phase and an undifferentiated grain ice mantle. The sticking probability is assumed to be unity when a neutral atom or molecule
collides with a grain surface, except for H and D atoms, for which we adopt the temperature-dependent sticking probability of \cite{tielens85}.
Adsorption energies of molecules on gain surfaces are generally taken from \cite{garrod06}.
The adsorption energies of deuterated species are generally set to be the same as those of normal species. One exception is  the D atom, the adsorption energy of which is set to be 21 K higher
than that of the H atom, which is assumed to be 600 K \citep{caselli02}. The adsorption energies of CO, N$_2$ and HCN, the abundance of which we present in the following section, are set to be
1150 K, 1000 K, and 3370 K, respectively. The value for HCN is adopted from the Temperature Programmed Desorption (TPD) experiment by \cite{noble13}.
The dependence of molecular abundances on the adsorption energies of CO and N$_2$ are presented in \cite{aikawa15}.

In addition to  thermal desorption, we take into account three non-thermal desorption processes: photodesorption, stochastic heating
by cosmic rays, and reactive desorption \citep[e.g.][]{oberg09a,hase93,garrod07}. The efficiency of chemical desorption is set to be
$10^{-4}-10^{-2}$ depending of the reactions, referring to \cite{garrod07}.
Recent laboratory experiments show that the efficiency depends on the molecular composition of the grain surface \citep{minissale16}. In order to include such an effect, we need to adopt the three-phase model, 
discriminating the ice surface from the bulk ice mantle, which we postpone to future work.
We assume that the surface reactions occur via the Langmuir-Hinshelwood mechanism; i.e. the adsorbed species diffuse on grain surfaces via thermal hopping, and react with each other when they meet, before they
desorb. The modified rate of \cite{caselli98} is adopted to ensure that the rate of hydrogenation (deuteration) is not higher than the adsorption rate of H atoms (D atoms)
when the number of such atoms on a grain is small ($\lesssim 1$).
The barrier for thermal hopping ($E_{\rm diff}$) is set to one-half of the adsorption energy ($E_{\rm ads}$). Lower values from 0.3 to 0.4 have also been used in astrochemical models \citep[e.g.][]{ruaud16, penteado17};
the surface reactions become more rapid with the lower ratio of $E_{\rm diff}/E_{\rm ads}$. The efficiency of H$_2$ formation and those of its isotopologs on granular surfaces become lower in warmer regions
(e.g. $T\gtrsim$ a few $10$ K) in the disk, since the physisorbed H (and D) atoms are desorbed more promptly. When the temperature is high enough, however, a bare grain surface appears, on which H atoms can
be chemisorbed. At such high temperatures, we assume that the formation rate of H$_2$ is 0.2 times the sticking rate of H atom onto grain surfaces, referring to \cite{cazaux04,cazaux10} \cite[see eq. 19 of][]{furuya13}.

The elemental abundance of deuterium is set to be $1.5\times 10^{-5}$ relative to hydrogen \citep{linsky03}.
The initial molecular abundances are determined by considering the molecular evolution from a cloud formation stage 
to a collapse phase to form a protostar. \cite{furuya15} first calculated molecular evolution behind the shock front of colliding HI gas. \cite{furuya16} then calculated the subsequent molecular evolution
in the hydrostatic prestellar core and in its collapse phase to form a protostar, as in \cite{aikawa12}. We adopt the fiducial model of \cite{furuya16} for our initial abundances; the shock model is run until the column density of the post shock gas
reaches a visual extinction of 2 mag, and the duration of the hydrostatic prestellar phase is set to be $10^6$ yr. Then the core collapses to form a protostar in $2.5\times 10^5$ yr, and the
molecular evolution in the infalling envelope is  calculated until the protostar is $9.3\times 10^4$ years old. While \cite{furuya16} calculated a spatial distribution of molecular abundances in a protostellar core,
our initial abundances for the present work are adopted from the infalling fluid parcel which reaches a radius
of 60 au in the core. The initial abundances of major species are presented in Table \ref{initial}; since the temperature ($\sim$ 144 K) and density ($\sim 5\times 10^7$ cm$^{-3}$) are high in the infalling fluid parcel at 60 au,
ices and ions are not abundant.
While the OPR of H$_2$ is initially $3.2 \times 10^{-3}$, it reaches thermal equilibrium or steady state in a relatively short timescale (see Appendix A).
Our results thus do not significantly depend on the initial OPR of H$_2$.

\section{RESULTS}

In the following, we consider the spatial distributions of HCO$^+$, DCO$^+$, N$_2$H$^+$, N$_2$D$^+$, HCN, and DCN, and describe the major formation pathways of the deuterated species.
We present the results at $3\times 10^5$ yr, which is earlier than a typical age of T Tauri stars ($\sim 1$ Myr), because we do not include  radial
accretion and/or mixing in the present model. The abundances of HCO$^+$ and N$_2$H$^+$ (and their isotopologs) are strongly coupled with
CO and N$_2$, which are gradually converted to less volatile species such as CO$_2$ and NH$_3$, and depleted from the gas phase.
The timescale of this conversion in the gas phase is, however,  $\sim 5\times 10^5 \left(\frac{\zeta_{\rm He}}{2.5\times 10^{-17} {\rm s}^{-1}}\right)^{-1}$ yr, where $\zeta_{\rm He}$ is the ionization rate of He atom.
It is comparable to or longer than the accretion timescale of the disk at $r\lesssim 100$ au \citep{furuya14}\citep[see also][]{furuya13}.
In other words, gaseous CO and N$_2$ can be supplied from the outer radius by viscous accretion and the radial drift of dust grains with ice mantles.
The abundances of CO and N$_2$ are thus underestimated, if we choose 1 Myr in our static models. The temporal variation of the abundance of each molecular species is also described in the following.

\subsection{Fiducial Model}

 \subsubsection{HCO$^+$ and DCO$^+$}
\begin{figure}
\includegraphics[scale=0.7, angle=90]{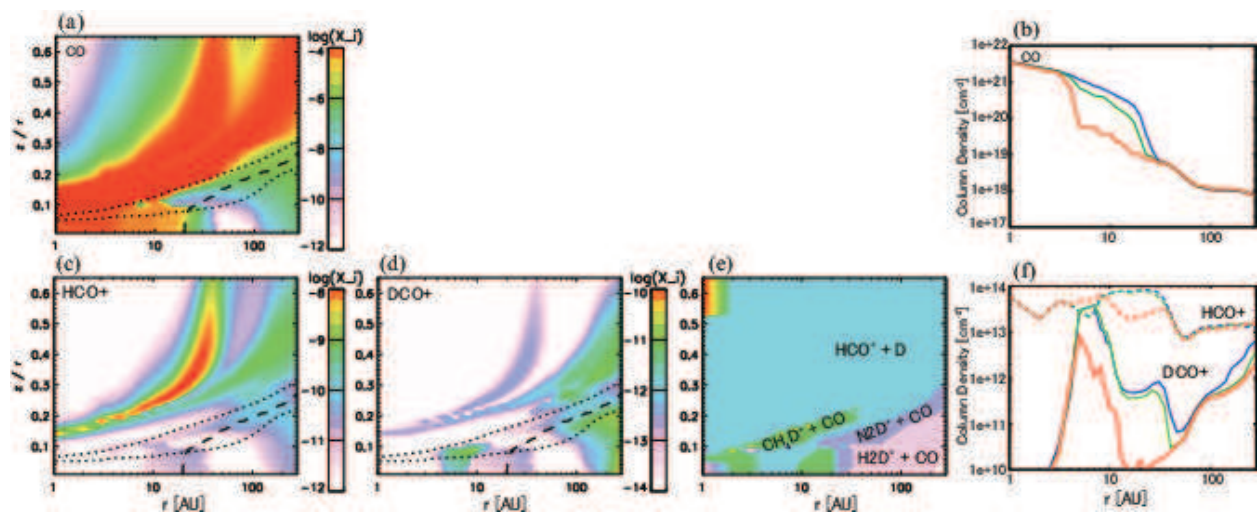}
\caption{Panels (a), (c), and (d) depict distributions of CO, HCO$^+$, and DCO$^+$ in the gas phase in the disk model with mm-sized grains at  $3\times 10^5$ yr. The long dashed lines depict the CO snow surface,
while the dotted lines are the same as in Figure \ref{dist_phys}. 
Panel (e) is color coded referring to the major formation pathways of DCO$^+$ at each position in the disk.
Panels (b) and (f) show the radial distribution of molecular column densities at $1\times 10^5$ yr (blue), $3\times 10^5$ yr (green), and $9.3\times 10^5$ yr (red). 
\label{dist_1mm_1}}
\end{figure}

Let us first consider HCO$^+$ and DCO$^{+}$.  The former is mainly formed by the reaction CO + H$_3^+$. As a reference, distributions of gaseous CO abundance and its column density are shown in
Figure \ref{dist_1mm_1} (a) (b), while those of HCO$^+$ and DCO$^+$ are shown in Figure \ref{dist_1mm_1}, panels (c)(d) and (f).
The horizontal axis depicts the radius ($r$) of the disk. In panels (a) (c) (d) and (e), the vertical axis is the distance from the midplane $z$, normalized by the radius; i.e. $z/r$, while the vertical axis
is the column density for panels (b) and (f). In the panels showing the abundance distributions (a, c and d), the dotted lines depict the positions where the X-ray ionization rate is equal to the cosmic-ray
ionization rate ($5\times 10^{-17}$ s$^{-1}$) (the upper dotted line) and to the ionization rate by decay of radio active nuclei ($1\times 10^{-18}$ s$^{-1}$) (the lower dotted line). The long dashed line
indicates the CO snow surface; it indicates the positions where CO abundances in the gas phase and in the ice mantle should be equal,
if the abundances are determined simply by adsorption onto grains and thermal desorption. In the layer below the long dashed line, CO is expected to be mainly in the ice mantle, while
it is expected to be abundant in the gas phase, elsewhere. The snow line is defined as the radius at which the snow surface crosses the disk midplane.

In our model, CO is gradually converted to less volatile molecules such as CO$_2$ ice and CH$_3$OH ice and thus depleted even in the regions inside the CO snow line ($\sim 20$ au) and in the layer above the CO snow surface \citep{furuya14}.
Although we present the results at $3\times 10^5$ yr, which is shorter than the timescale of chemical conversion in the
gas phase, chemical conversion on grain surfaces is more efficient than in the gas phase.
In the midplane at 40 au $\lesssim r \lesssim 100$ au, CO (both in the gas phase and in ice mantle), and thus HCO$^+$, are deficient ($n(i)/n({\rm H})\lesssim 10^{-11}$), and such regions
with low CO and HCO$^+$ abundances extend inwards over time.
For example, the gaseous CO abundance is as low as $5\times 10^{-9}$ and the HCO$^+$ abundance is $1\times 10^{-12}$ in the midplane at $r=10$ au at $t=1$ Myr. 
It should be noted that ALMA observations recently revealed such depletion of gaseous CO inside the CO snow line in TW Hya \citep{nomura16, schwarz16} \citep[see also][]{zhang17}.
In the midplane at inner radii ($r\lesssim 10$ au), the HCO$^+$ abundance is limited by the low ionization degree, which occurs at high density.
At $r\lesssim 70$ au, molecular ions such as H$_3^+$ and HCO$^+$ are deficient at $z/r\sim0.15$, because
photoionization makes atomic ions (e.g. S$^+$) the dominant positive charge carrier in such upper layers. The high electron abundance then reacts quickly with molecular ions.

Figure \ref{dist_1mm_1} (e) depicts the major formation pathways of DCO$^+$ at each position in the disk. Comparing panels (d) and (e), we can see that
DCO$^+$ is abundant in the layer below the CO snow surface (i.e. at the lower $z$) due to  reaction (\ref{dcop}), despite the relatively low CO abundance in the gas phase.
In the midplane, deuterated-H$_3^+$ increases towards the outer radius, where the temperature is lower and the chemical conversion of HD is less efficient  (see Appendix B).
The chemical conversion of CO is also less efficient at outer radii ($\gtrsim 200$ au). The DCO$^+$ abundance thus increases
outwards at $r\gtrsim 100$ AU. In the upper layers ($z/r\gtrsim 0.2$), on the other hand, DCO$^+$ is mainly formed via
\begin{equation}
{\rm HCO}^+ + {\rm D} \rightarrow {\rm DCO}^+ + {\rm H} + 796 {\rm K}. \label{hcopd}
\end{equation}
Since the reaction is exothermic by 796 K \citep{adams85}, the backward reaction is less efficient than the forward reaction even at warm temperatures.

\begin{figure}
\includegraphics[scale=0.7, angle=0]{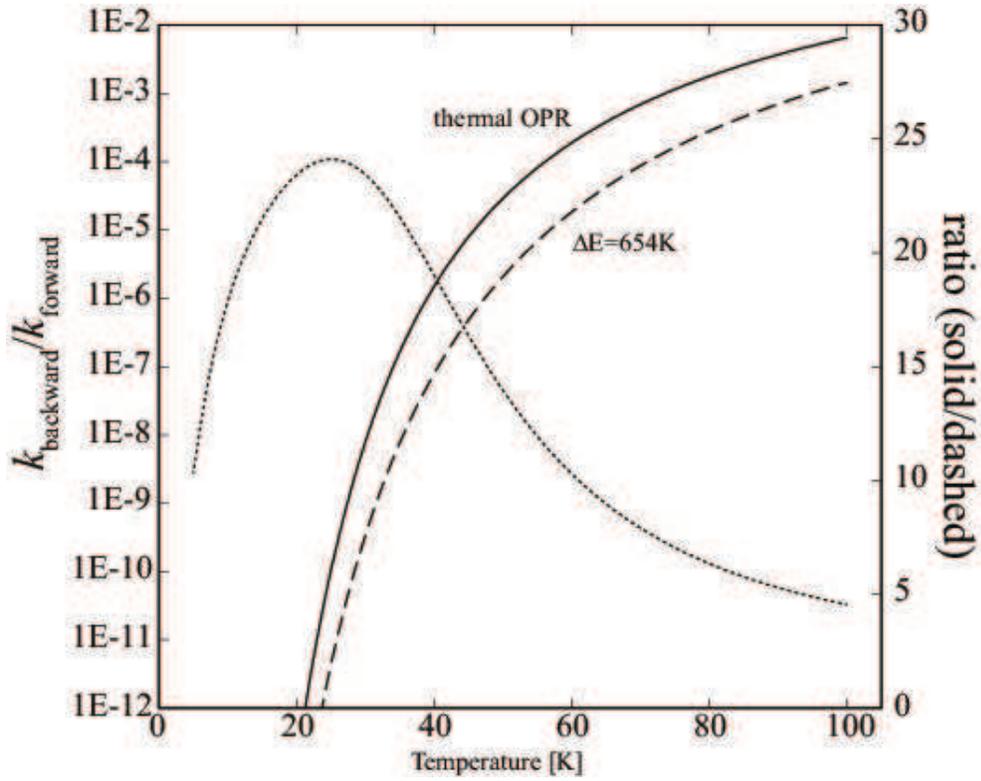}
\caption{The ratio of  backward to forward reaction rate coefficients of CH$_3^+$ + HD $\rightarrow$ CH$_2$D$^+$ + H$_2$  (reaction \ref{ch2dp}) as a function of gas temperature given in \cite{roueff13}.
The solid line depicts the value with the thermal OPR of H$_2$, while the dashed line depicts the value with a constant exothermicity of 654 K assuming all H$_2$ in the para form. The ratio of the former to the latter is shown by the dotted line.
\label{rate}}
\end{figure}

These results are basically the same as described in \cite{oberg15}. One update is that, following \citet{roueff13}
we set the exothermicities of reaction (\ref{ch2dp}) and its multi-deuterated counterparts to values higher than those previously determined \citep{smith82} and
used in \cite{oberg15}. For example, the exothermicity of CH$_3^+$ + p$\mathchar`-$H$_2$ and CH$_3^+$ + o$\mathchar`-$H$_2$ are set to  660 K and 489 K, respectively.
Although we do not distinguish the nuclear spin state of CH$_3^+$, this omission does not affect our results, since the difference in exothermicities among the reactions with ortho- and
para-CH$_3^+$ is only 6 K \citep{roueff13}.
Although DCO$^+$ is also formed via the reaction between CO and CH$_4$D$^+$, where the ion is produced via reaction (\ref{ch2dp}) and subsequent radiative association with H$_2$,
this path is the major formation path of DCO$^+$ only in limited regions, i.e. the green regions in panel (e).
It is in contrast to the model of \cite{favre15}, in which reaction  (\ref{ch2dp}) significantly contributes to the DCO$^+$ formation in warm surface layers and thus enhances the DCO$^+$ column density.  
Although the direct comparison between our model results and those of \cite{favre15}
is not straightforward due to the differences in disk physical structure, knowledge of the spin state of H$_2$ would be a key to differentiate between the two results. 
While \cite{favre15} assumed all H$_2$ in the para form, and used a constant exothermicity of 654 K for reaction (\ref{ch2dp}), we found that the OPR of H$_2$ is almost thermal
at each position in the disk (see Appendix A); the effective exothermicity of reaction (\ref{ch2dp}) thus decreases as o-H$_2$ increases with temperature.
Figure \ref{rate} shows the ratio of the backward to forward reaction rate coefficients referring to \cite{roueff13}.
For the solid line, we assumed that the OPR of H$_2$ is thermal. The dashed line, on the other hand, shows the ratio when all H$_2$ is in para state, so that the exothermicity of the forward
reaction is constant at 654 K. The dotted line depicts the ratio of the former (i.e. the ratio shown with the solid line) to the latter (the dashed line); it reaches a maximum at the temperature
of $\sim 30$ K,  which seems to correspond to the layer with abundant DCO$^+$ claimed by \cite{favre15}. 
It should be noted, however, that the contributions of reaction (\ref{ch2dp}) become more significant when the elemental C/O ratio is higher than unity (\S 3.2).

Figure \ref{dist_1mm_1} (f)  shows the radial distributions of column densities of HCO$^+$ (dashed lines) and DCO$^+$ (solid lines).
The blue, green, and red lines depict the values at $t=1\times 10^5$ yr, $3\times 10^5$ yr (the fiducial value), and $9.3 \times 10^5$ yr, respectively.
Both HCO$^+$ and DCO$^+$ increase inwards from $r\sim 40$ au, which is slightly outside the CO snow line ($\sim 20$ AU).  
In the midplane inside $r \sim 40$ au, thermal desorption of CO becomes non-negligible, which makes HCO$^+$ the dominant charge carrier,
while H$_3^+$ and its isotoplogues dominate in the outer radius \citep[see eq. (23) in][]{aikawa15}.
Inside a radius of $\sim 4$ au, the DCO$^+$ abundance is low in the midplane; the warm temperature lowers the atomic D/H ratio, and the D atom abundance is also lowered
by  reaction with HS. Outside 50 au, the column density of DCO$^+$ increases outwards due to its enhancement in the outer midplane, while the HCO$^+$ column density is relatively flat. 
Both column densities gradually decrease with time, as CO is converted to less volatile species. The decline of the DCO$^+$ column density is more significant than that of HCO$^+$, 
since DCO$^+$  has its abundance peak in the midplane at  outer radii, where CO conversion is efficient.

\subsubsection{N$_2$H$^+$ and N$_2$D$^+$}
\begin{figure}
\includegraphics[scale=0.7, angle=90]{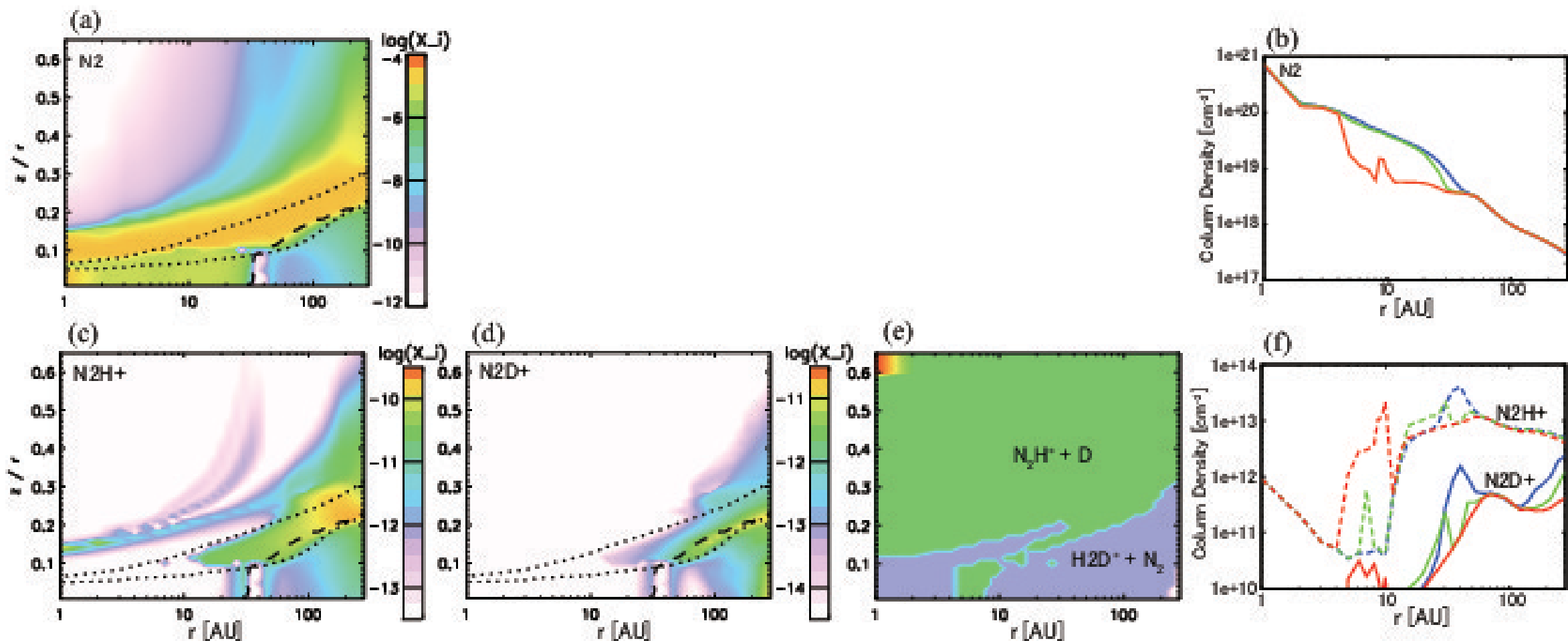}
\caption{Distributions of molecular abundances (a, c, and d) and molecular column densities (b and f) as in Figure \ref{dist_1mm_1}, but for
N$_2$, N$_2$H$^+$, and N$_2$D$^+$ in the gas phase.
The long dashed lines in panels (a), (c) and (d) depict the N$_2$ snow surface. Panel (e)  is color coded referring to the major formation pathways of N$_2$D$^+$ at each position in the disk.
\label{dist_1mm_2}}
\end{figure}

Figure \ref{dist_1mm_2} shows distributions of molecular abundances, column densities, and major formation pathways of deuterium isotopologs as in Figure \ref{dist_1mm_1}, but for N$_2$,
N$_2$H$^+$, and N$_2$D$^+$.
N$_2$H$^+$ is formed by N$_2$ + H$_3^+$, and is destroyed by recombination with electrons and proton transfer to CO. Thus
its abundance depends on N$_2$, CO, H$_3^+$ and electrons \citep{aikawa15}; the abundance has a peak
around the CO snow surface, where the abundance ratio of CO to electron is $\sim 10^3$, and in the layer above the CO snow surface,
where gaseous CO is converted to less volatile species to be frozen onto grains.

In the midplane, N$_2$ is gradually converted to less volatile species such as  NH$_3$ ice, and thus the region of low N$_2$H$^+$ abundance basically expands with time,
although N$_2$H$^+$ becomes abundant at $t=9.3\times 10^5$ yr around the radius of several au, where gaseous CO is reduced by the conversion effect.
 The formation pathways of N$_2$D$^+$ are similar to those of DCO$^+$; N$_2$D$^+$ in the midplane is formed by the reaction of N$_2$ with deuterated H$_3^+$,
while N$_2$D$^+$ is formed by 
\begin{equation}
{\rm N}_2{\rm H}^+ + {\rm D} \rightarrow {\rm N}_2{\rm D}^+ + {\rm H} + 550 {\rm K} \label{n2hpd}
\end{equation}
above the CO snow surface. In the upper layers, the ratio of N$_2$D$^+$/N$_2$H$^+$ is lower than that of DCO$^+$/HCO$^+$, because the exothermicity of reaction (\ref{n2hpd}) (550 K) is lower than that of reaction (\ref{hcopd}).
The column densities of both N$_2$H$^+$ and N$_2$D$^+$ have a peak at 20 au $\lesssim r \lesssim$ 50 au. The inner boundary corresponds to the snow line of CO, the main reactant with N$_2$H$^+$, while the outer boundary
is slightly outside the N$_2$ snow line, for a similar reason used for HCO$^+$. Outside $r\sim 150$ au, the N$_2$D$^+$ column density increases outwards, since it is abundant in the outer midplane.

\subsubsection{HCN and DCN}

\begin{figure}
\includegraphics[scale=0.7, angle=90]{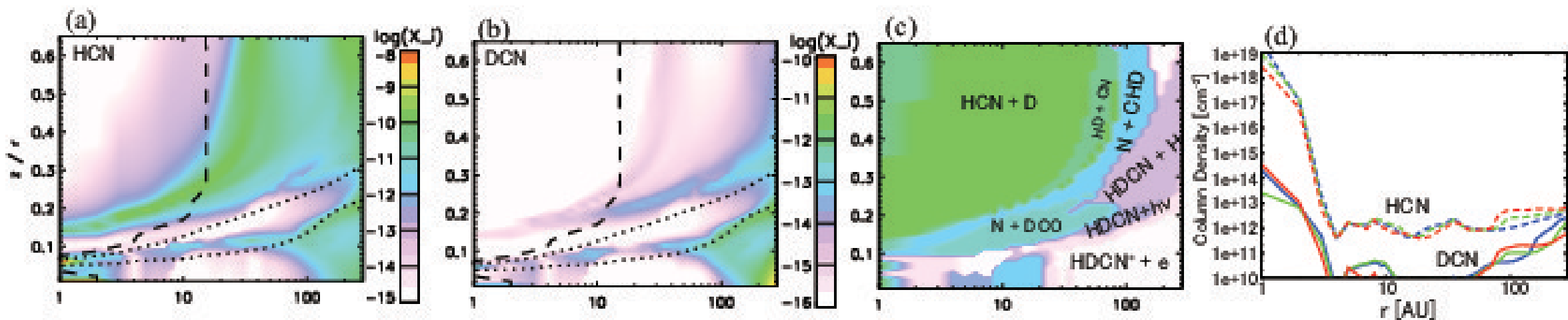}
\caption{Distribution of molecular abundances (a-b) and molecular column densities (d) as in Figure \ref{dist_1mm_1}, but for
HCN, and DCN in the gas phase.
The long dashed lines in panels (a) and (b) depict the HCN snow surface. Panel (c)  is color coded referring to the major formation pathways of DCN at each position in the disk.
\label{dist_1mm_3}}
\end{figure}

Figure \ref{dist_1mm_3} shows the distributions of HCN and DCN,  the formation pathways of DCN, and the column densities of HCN and DCN. Since the desorption energy of HCN is high (3370 K), the abundance
of gaseous HCN is $\lesssim 10^{-10}$ outside a radius of a few au, where the temperature is $\lesssim 100$ K. HCN in the ice mantle, on the other hand, is as abundant as $10^{-7}-10^{-6}$ in the midplane.
Towards the outer ($r\gtrsim 30$ au) midplane, HCN forms by the dissociative recombination of H$_2$CN$^+$, which forms by H$_3^+$ + CN $\rightarrow$ HCN$^+$ + H$_2$ followed by
HCN$^+$ + H$_2$ $\rightarrow$ H$_2$CN$^+$ + H. In the upper layers and inner radii, it forms by H + H$_2$CN and N + HCO. The precursor molecules
H$_2$CN and HCO are produced by CH$_3$ + N and CH$_2$ + O, respectively. HCN also is produced by reactive desorption  after the grain surface reaction of H + CN. 

The major formation paths of DCN are mostly the deuterated version of the above reactions. One exception is 
\begin{equation}
{\rm HCN + D \rightarrow DCN + H,} \label{schilke}
\end{equation}
which is effective at the disk surface, depicted by the green region in Figure \ref{dist_1mm_3} (c). In our model, the activation barrier of both the forward and backward reaction of (\ref{schilke})
is set to be 500 K, following \cite{schilke92}.
A more recent quantum chemical calculation (Kayanuma in private communication) evaluates this barrier to be 3271 K, but also predicts that the effect of the barrier would be significantly lowered by tunneling. 
Our result does not change, even if we assume a higher barrier, e.g. 1800 K, for reaction (\ref{schilke}).
Another exception is C$_3$H$_4$D$^+$ + N, which dominates in the midplane at $r\sim 6$ au, i.e. the blue region.
In this warm midplane region (40 K $< T < $ 50 K), hydrocarbons are also deuterated via 
\begin{equation}
{\rm C_2H_2^+ + HD \rightarrow C_2HD^+ + H_2,} \label{c2h2p}
\end{equation}
which is exothermic by 550 K.
The major exothermic exchange reaction that initiates the enhancement of the DCN/HCN ratio is thus reaction (\ref{h2dp}) in the outer midplane, while reactions (\ref{ch2dp}) and (\ref{c2h2p}) dominate
in the upper layers and in inner radii.
It should be noted that HCND$^+$ and  DCNH$^+$ are not distinguished in our reaction network, which means that the network of DCN and DNC is partially mixed in regions where the recombination
of HDCN$^+$ is their major formation path. 

Due to the contribution from the outer midplane, the column density of DCN increases outwards at $r\gtrsim 20$ au.
Inside a radius of a few au, both the HCN and DCN column densities are high, since they are relatively abundant in the $z/r\sim 0.06$ and $z/r \sim 0$ layers. In the upper layer, large hydrocarbons in ice mantles such as H$_5$C$_3$N
serve as reservoirs of carbon and nitrogen, while the reactants of C-bearing species, such as O atoms, are deficient in the midplane with high density.

\subsection{Elemental Abundances}
Observations in recent years suggest that the surface and molecular layers are deficient in elemental carbon and oxygen, especially in relatively cold disks.
\cite{hogerheijde11} detected ground-state rotational emission lines of H$_2$O in the disk around TW Hya using the Herschel Space Observatory to find that the emissions are significantly weaker than expected
from disk models. \cite{du17} obtained and analyzed the water emission of 13 protoplanetary disks. They compared the observational data with disk models to show that the abundance of gas-phase oxygen needs to be
reduced by a factor of at least $\sim 100$ to be consistent with the observational upper limits and positive detections, if a dust-to-gas mass ratio is 0.01.
\cite{meijerink09} compared the radiative transfer models with mid-infrared spectrum of H$_2$O taken by {\it Spitzer} to find that water vapor is significantly depleted
in the disk surface beyond the radius of $\sim 1$ au. They proposed that water vapor is depleted by the vertical cold finger effect; turbulent diffusion transports the water vapor
from the disk surface to the layer below the snow surface, where water can freeze out and transported to the midplane via dust settling.
\cite{favre13} found that the gaseous CO abundance is low ($10^{-6}-10^{-5}$) even  in the warm molecular layer (above the CO snow surface) in the disk of TW Hya by comparing C$^{18}$O ($J=2-1$) and HD ($J=1-0$)
emission line intensities. \cite{kama16} combined various emission lines including CI and OI towards TW Hya and HD100546; carbon and oxygen were found to be strongly depleted from the gas phase in the disk of TW Hya, while the depletion
is moderate in the disk of HD100546. \cite{kama16} also presented an analytical model of the vertical cold finger effect to show that a combination of turbulent mixing and settling of large dust grains deplete C- and O-bearing volatiles
from the surface and molecular layers of disks by locking them in the ices in the midplane \citep[see also][]{krijt16, xu17}.
Since H$_2$O is less volatile than CO, such a depletion mechanism would be more efficient for oxygen than carbon, which could enhance the C/O ratio in the surface and molecular layers.
\cite{kama16} found the C/O ratio to be higher than unity in the disk of TW Hya. \cite{bergin16} observed C$_2$H emission, which is bright in a ring region, in TW Hya and DM Tau. In order to reproduce the bright
C$_2$H emission in disk models, a high C/O ratio ($>1$) is required together with a strong UV field.

\begin{figure}
\includegraphics[scale=0.7, angle=90]{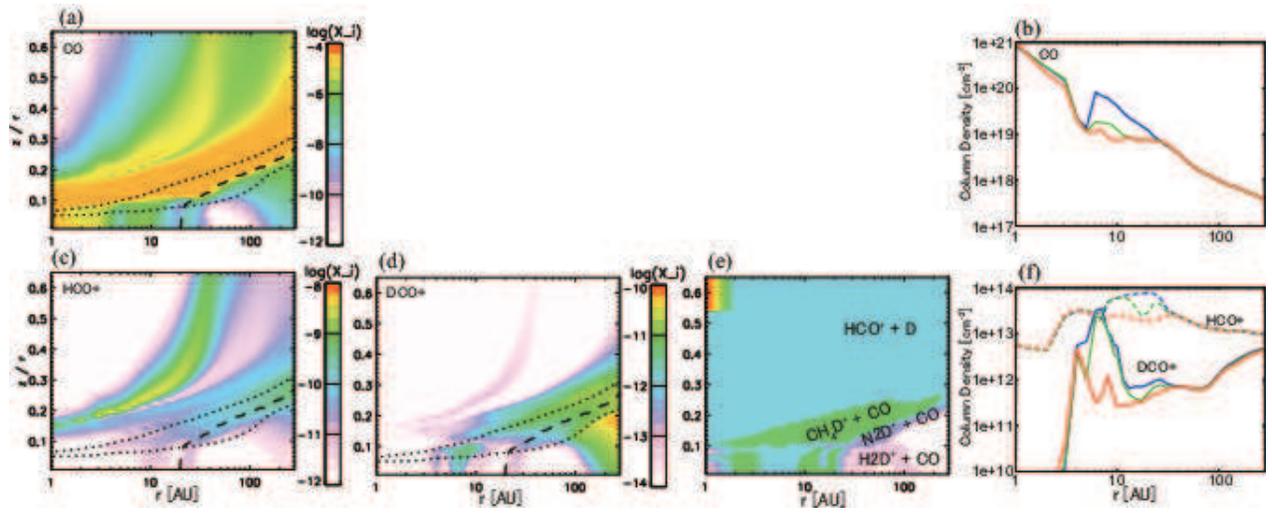}
\caption{Same as Figure \ref{dist_1mm_1}, but for the disk model with depletions of CO and H$_2$O. In panel (e), DCO$^+$ is formed mainly via H$_2$D$^+$ + CO in the pink region, while
the reactions of multi-deuterated H$_3^+$ with CO dominate in the white region.
\label{dist_COdep_1}}
\end{figure}

In order to investigate the depletion of elemental carbon and oxygen on deuterium chemistry, we performed a calculation of our fiducial disk model as in \S 3.1, but with a modified set of  initial abundances.
We set the initial abundance of H$_2$O to zero and reduce the CO abundance by an order of magnitude, while the abundances of other species are the same as in our fiducial model.
Even with this reduced abundance, CO is still the major carbon carrier, although CH$_4$ is the most abundant among C-bearing species (see Table 1). Major oxygen carrier in the initial condition is
CO, H$_2$CO, and CO$_2$. The elemental C/O ratio is 1.43.
Figure \ref{dist_COdep_1} shows the distributions of CO, HCO$^+$ and DCO$^+$ abundances (panels a, c, and d), their column densities (panels b and f) and major formation pathways of DCO$^+$ (panel e).
While the spatial distributions of the molecular abundances and column densities of HCO$^+$ and DCO$^+$ are basically similar to the fiducial model (Figure \ref{dist_1mm_1}), there are several notable differences.
Firstly, CO depletion via chemical conversion to CO$_2$ ice is less effective above the CO snow surface in the outer radius ($r\gtrsim 20$ au) than in the fiducial model, which is natural considering the reduced
oxygen abundance. HCO$^+$ and DCO$^+$ thus becomes more abundant in the layer above the CO snow surface, in which DCO$^+$ is mainly formed by reaction (\ref{ch2dp}).
The high C/O ratio enhances the abundance of hydrocarbons and thus the importance of reaction  (\ref{ch2dp}).
 In spite of the reduced CO abundance, the column densities of HCO$^+$ and
DCO$^+$ are similar to those in the fiducial model, except that the depression of the DCO$^+$ column density at $r\sim 50$ au is more modest and the HCO$^+$ column density is reduced at the central region ($r \lesssim$
a few au), in which HCO$^+$ exists mostly in the surface layer. 

\begin{figure}
\includegraphics[scale=0.7, angle=90]{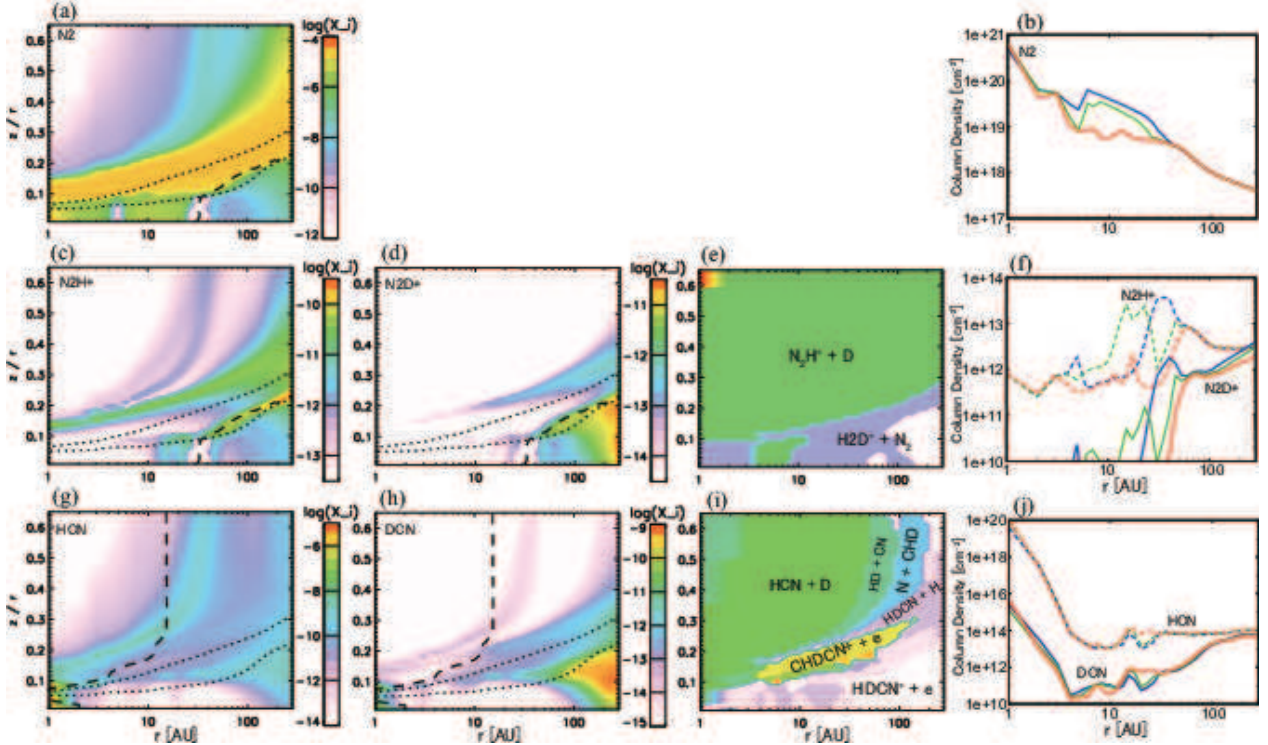}
\caption{Distribution of molecular abundances (a, c, d, g and h), and molecular column densities (b, f, and j) as in Figure \ref{dist_1mm_1}, but for
N$_2$, N$_2$H$^+$, N$_2$D$^+$, HCN and DCN in the gas phase in the disk model with depletions of CO and H$_2$O.
The long dashed lines depict the snow surface of the mother molecule. Panels (e) and (i) are color coded referring to the major formation pathways of N$_2$D$^+$ and DCN, respectively,
at each position in the disk. In panel (e), the reactions of multi-deuterated H$_3^+$ with N$_2$ dominate in the white region, while the reaction of H$_2$D$^+$ + N$_2$ dominates in the purple region.
\label{dist_COdep_2}}
\end{figure}

Figure \ref{dist_COdep_2} (a-f) shows the distributions of N$_2$, N$_2$H$^+$, and N$_2$D$^+$, their column densities, and the major formation pathways of N$_2$D$^+$. The distributions of N$_2$ and N$_2$D$^+$ are
similar to those in our fiducial model (Figure \ref{dist_1mm_2}), while N$_2$H$^+$ is redued in the layer above the CO snow surface, in which the gaseous CO abundance exceeds that in the fiducial model.

Figure \ref{dist_COdep_2} (g-j) shows the distributions of HCN and DCN, their column densities, and the major formation pathways of DCN.  Their abundances and column densities are significantly higher than in
the fiducial model, since the high C/O ratio enhances the abundance of hydrocarbons, which react with N atoms to form HCN.

We also calculated a model in which the initial abundance of H$_2$O is totally depleted but CO is not (the C/O ratio then becomes 1.10); the results are quite similar to the model described above,
except that CO is more abundant inside the CO snow line, which lowers the N$_2$H$^+$ abundance .

\subsection{Dust Grain Sizes}
Figure \ref{dist_ism_1} shows distributions of gaseous CO, HCO$^+$, and DCO$^+$ (panels a, c and d), the main formation paths of DCO$^+$ (panel e), and their column densities (panels b and f)
in the model with dark cloud dust. The initial abundances are the same as in our fiducial model.
Since the total surface area of grains is larger than in the fiducial model, the freeze-out of CO
and subsequent conversion to other molecules is more efficient in the dark cloud dust model than in the mm-sized grain model, especially
in the midplane regions. In the layer above the midplane, on the other hand, UV radiation is more efficiently attenuated, so that the molecular layer extends to larger $z$ than in Figure \ref{dist_1mm_1}.

At early time (e.g. $t=1\times 10^5$ yr), HCO$^+$ is abundant in the layer above the CO snow surface, and in the midplane inside the CO snow line. Since the contribution from the upper layers is
significant, the radial distribution of the HCO$^+$ column density is rather flat; although it slightly increases inwards around the CO snow line ($r\sim 100$ au), the increment is not significant compared with
that in Figure \ref{dist_1mm_1} at $r\sim 40$ au. HCO$^+$ in the midplane decreases with time, as CO is converted to less volatile species. Its column density, however, 
varies by less than a factor of three, due to its constantly high abundance in the upper ($z/r\gtrsim 0.3$) layers.

The DCO$^+$ ion is mainly formed by  reaction (\ref{hcopd}), since CO is depleted in the midplane, where reaction (\ref{dcop}) should be efficient. At the early time, D atoms are abundant
in the upper ($z/r\gtrsim 0.2$) warm layers, where reformation of HD is inefficient. DCO$^+$ at that stage is thus abundant in the upper layers, and its column density distribution is relatively flat
at $r\gtrsim 30$ au, including the radius around the CO snow line.
Then DCO$^+$ decreases as D atoms are incorporated into hydrocarbons, and CO inside the snow line is converted to less volatiles species. Its column density thus decreases significantly with time.
Even at $9.3 \times 10^5$ yr, DCO$^+$ in the midplane does not significantly contribute to its column density except around a radius of a few tens of au, where the column density has a sharp peak.
This is in contrast to the model with mm dust grains, in which DCO$^+$ in the midplane mostly determines its column density distribution.

In order to check the significance of the updated exothermicity of reaction (\ref{ch2dp}), we also ran a model in which its exothermicity is set to 370 K. 
Compared with the model with this old value, the DCO$^+$ abundance is enhanced at  $0.3 \lesssim z/r \lesssim 0.4$ at a radius of several tens of au in the present model, although
reaction (\ref{hcopd}) is the dominant formation path  for DCO$^+$ there.

Figure \ref{dist_ism_2} shows the distributions of N$_2$, N$_2$H$^+$, HCN and their deuterated isotopologs, their column densities, and the major formation pathways of the deuterated isotopologs in the
model with dark cloud dust.
N$_2$H$^+$ is abundant in the upper layers, where H$_3^+$ is the dominant ion, and in the midplane region between the snow lines of CO ($\sim 120$ au) and N$_2$ ($\sim 230$ au). Its abundance in the midplane, however,
decreases as N$_2$ is converted to NH$_3$ ice in a few $\times$ $10^5$ yr \citep{furuya14}. The distribution of N$_2$D$^+$ is similar to that of N$_2$H$^+$, but its abundance in the upper ($z/r\gtrsim 0.3$) layer is
lower than the maximum abundance in the midplane. The column density of N$_2$D$^+$, therefore, decreases significantly as N$_2$ is converted to NH$_3$ ice in the midplane.
The major deuteration path is via H$_2$D$^+$ in the midplane, and via D atoms (reaction \ref{n2hpd}) in the upper layers.

Outside a radius of $r\sim 10$ au, gaseous HCN and DCN are distributed mostly in the upper ($z/r\gtrsim 0.2$) layers, where they form
via the recombination of H$_2$CN$^+$ (HDCN$^+$) and reaction H$_2$CN (HDCN) + H.
The radial distributions of their column densities gradually increase outwards at $r\gtrsim 10$ au. At the innermost radii ($r\lesssim 3$ AU), on the other hand, 
the temperature is high enough ($\gtrsim 100$ K) to desorb HCN and DCN from  ice mantles.

We also calculated molecular abundances in the disk model with the maximum grain (pebble) size of 10 cm, and found that the distributions of gaseous molecular abundances are qualitatively similar to those in our fiducial model (i.e. the model with mm-sized dust). Some notable differences are as follows. Firstly, the midplane temperature is slightly higher in the model with cm dust.
Secondly, CO is converted to CO$_2$ ice more efficiently in the model with cm dust inside the CO snow line (around a radius of a few tens of au), because the OH radical, which reacts with CO
to produce CO$_2$, is produced from H$_2$O ice due to the higher UV flux.
Thirdly, gaseous HCN and DCN are more abundant in the model with cm dust outside a radius of a few au; a deeper penetration of UV and slightly warmer temperature enhance their precursors, 
N atoms and hydrocarbons, in the gas phase. Inside their snow line ($\lesssim 2$ au), on the other hand, HCN and DCN are less abundant in the model with cm dust,
since more refractory carbon chains accumulate in the ice mantle.

\subsection{Turbulent Mixing} 

Protoplanetary disks are considered to be turbulent, most probably due to  magneto-rotational instability \citep{balbus91}. The direct measurement of non-thermal velocity dispersion $v$
has been one of the major challenges in radio observations of disks. The dispersion is basically subsonic with a Mach number $\mathcal{M}=v/c_s \sim 0.2-0.4$ \citep{teague16}, where
$c_s$ is the sound speed. While a very low velocity
dispersion $\mathcal{M}<0.03$ is derived in HD163296 \citep{flaherty15}, \cite{teague16} took into account the uncertainties in flux calibration to derive an  upper limit of $\mathcal{M}\sim0.16$.
Since the disk has vertical temperature gradients, and since the mixing timescale in the vertical direction is shorter than that in the radial direction \citep[e.g.][]{aikawa96}, the vertical mixing could alter the molecular D/H
ratios \citep{furuya13, albertsson14}. In this subsection, we investigate the effect of vertical turbulent mixing on chemistry including the D/H ratios.

The diffusion coefficient is of the same order as the kinematic viscosity coefficient, $\alpha c_s H$, where $H$ is the scale height of the disk.
Although there could be a slight difference between the two values \cite[e.g.][]{johansen05}, we assume that the values of the two coefficients are the same in the present work.
The non dimensional parameter $\alpha$ is equal to $(v/c_s)(l/H)\approx(v/c_s)^2$, where $l$ is the size of the turbulent eddy. The value of $\alpha$ is thus estimated to be $\lesssim 10^{-2}$
in protoplanetary disks. Figure \ref{diff_neutral} and Figure \ref{diff_ion} show distributions of neutral and ionic species, respectively, in models with a diffusion coefficient $\alpha= 10^{-3}$ (top row) and $10^{-2}$ (middle row).
The bottom panels show the
molecular column densities for the model without diffusion (solid lines) and with a diffusion coefficient of $\alpha=10^{-3}$ (dashed) and $10^{-2}$ (dotted).
CO and N$_2$ are also shown in Figure \ref{diff_neutral} in addition to HCN and DCN, since DCO$^+$ and N$_2$D$^+$ are chemically coupled to these precursor species.

The effect of turbulence is twofold. First, it transports ices to the disk surface, where the ices are thermally desorbed and photodissociated. Secondly, it transports H atoms and
radicals from the disk surface to the midplane, where they contribute to  grain-surface reactions.  In the static model, CO is efficiently converted to CO$_2$ via
the grain-surface reaction of CO + OH in the layer above the CO snow surface (\S 3.1). In the model with diffusion, on the other hand, CO is more abundant in this layer, 
since CO ice in the midplane is transported to this layer and sublimated, while CO$_2$ ice is transported to the disk surface to be photodissociated. 
Thus the column density of gaseous CO is slightly higher in the models with diffusion at $r\gtrsim 60 $ au.
In the midplane at 10 au $\lesssim r \lesssim 50$ au, on the other hand, CO is more efficiently converted
to CO$_2$ ($\alpha=10^{-3}$) and CH$_3$OH ($\alpha=10^{-2}$) in the models with diffusion, due to the enhanced abundances of OH and H atoms.
Inside a radius of $\sim 40$ au, HCO$^+$ is mostly on the disk surface, and is increased by the diffusion of CO in the turbulent disk.

The major nitrogen carriers are N$_2$ and NH$_3$ in our models.
In the models with diffusion, the N$_2$ column density is reduced at $r\gtrsim 50$ AU, while  it is enhanced at smaller radii, compared with the model without diffusion.
In the cold outer midplane region, NH$_3$ is more
efficiently formed via H atoms coming from the disk surface, while in the inner warm regions, midplane temperatures are too warm to (re)form NH$_3$ via grain-surface hydrogenation, and NH$_3$ is transported to the
disk surface to be photodissociated \citep{furuya14}. The high abundance of N$_2$ in the midplane enhances the abundance of N$_2$H$^+$ at 10 au $\lesssim r \lesssim 50$ au.

Diffusion of CO ice from the midplane and of H atoms from the disk surface enhance the abundances of hydrocarbons around and below the CO snow surface, and thus
HCN and DCN, as well. Inside a radius of a few au, HCN and DCN abundances in the midplane are significantly reduced by the vertical mixing.
Turbulence transports the large hydrocarbons, which represent the major carbon reservoir near the midplane of the static model, to the disk surface, and O atoms to the midplane, which destroys C-bearing species.


The vertical diffusion also affects the deuterim chemistry. First, it makes the depletion of HD in the midplane and conversion to HDO and NH$_2$D ices (see Appendix B) less efficient via the transport of the deuterated ices to the disk
surface, where they can be dissociated. Thus in the model with mixing, HD and D atoms are more abundant in the midplane at $r\gtrsim 40$ au.
Secondly, D atoms are transported towards the midplane and enhance deuteration via reactions such as (\ref{hcopd}).

\subsection{Cosmic-ray Ionization}

So far we have assumed that cosmic rays provide a minimum ionization rate of $5\times 10^{-17}$ s$^{-1}$ in the midplane, where X-ray penetration is significantly attenuated.
But the cosmic rays may be excluded by stellar winds with magnetic fields \citep[e.g.][]{umebayashi81,cleeves13a}. 
In order to check the effect of the exclusion of cosmic rays, we investigated the model without cosmic-ray ionization.
Here, the model disk is ionized by stellar X-rays, which dominate in and above the molecular layer, and the decay of radioactive nuclei, which sets a minimum ionization rate of $1\times 10^{-18}$ s$^{-1}$ 
in the midplane (below the lower dotted line in Figure \ref{cr18}), given
the $^{26}$Al abundance derived from the analysis of meteorites for the formation stage of the Solar System \citep{umebayashi09}. Although the ionization rate could be smaller depending on the abundances of 
radioactive nuclei and the surface density of the disk \citep{umebayashi09, cleeves13b}, the effect
of a low ionization rate is already apparent in the model presented here, and it is straightforward to extrapolate the results to an even lower ionization rate, at least qualitatively.

\begin{figure}
\includegraphics[scale=0.7, angle=90]{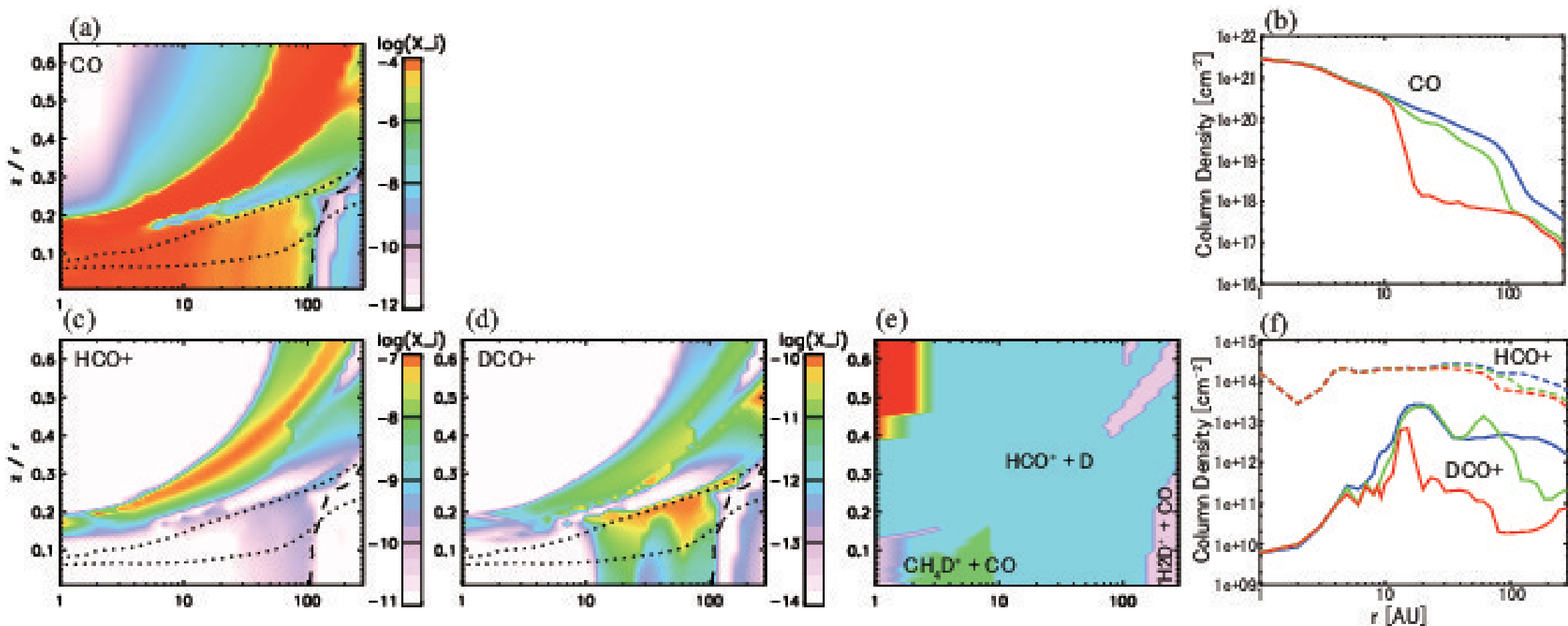}
\caption{Same as Figure \ref{dist_1mm_1}, but for the disk model with dark cloud dust.
\label{dist_ism_1}}
\end{figure}

\begin{figure}
\includegraphics[scale=0.7, angle=90]{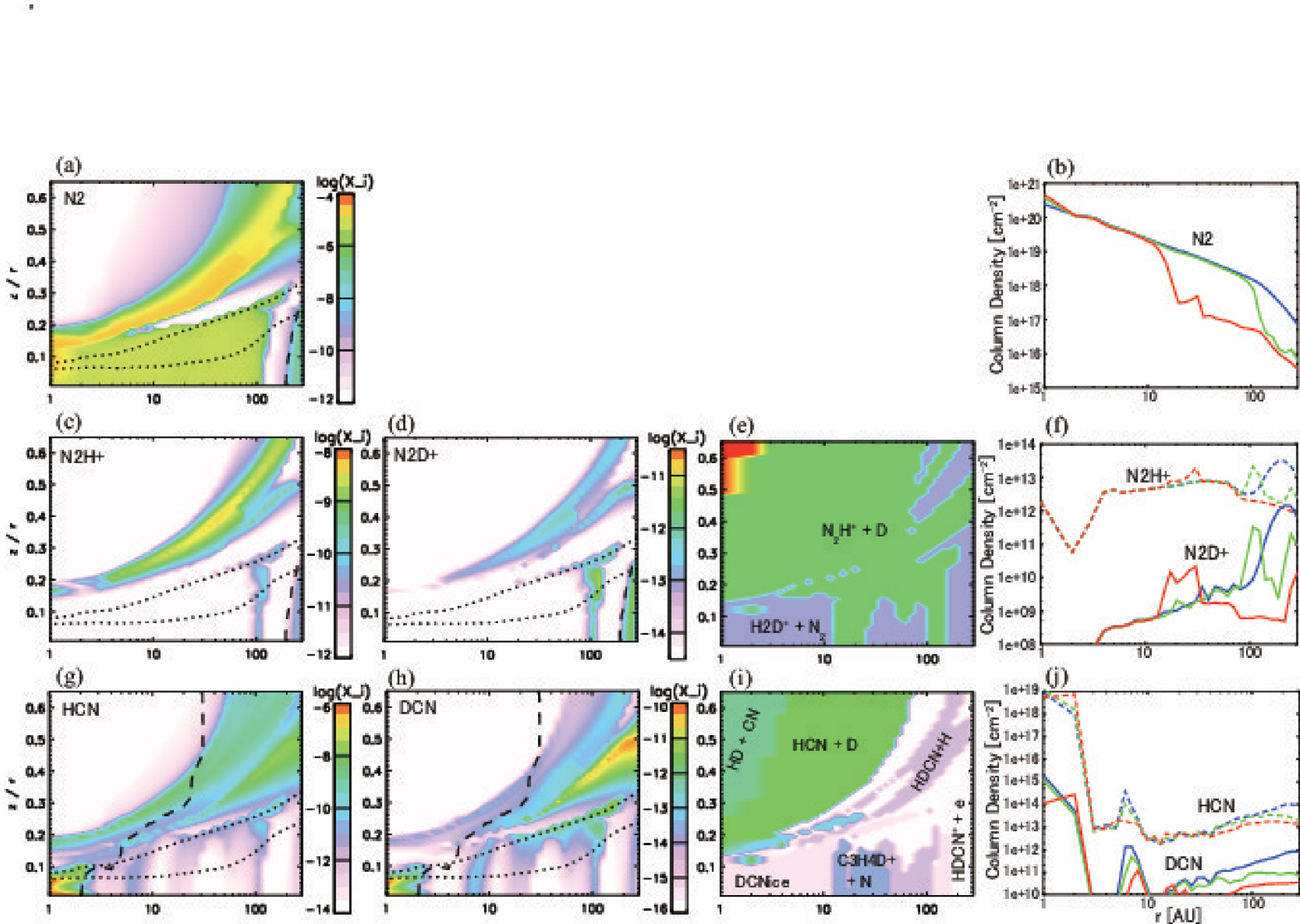}
\caption{Same as Figure \ref{dist_COdep_2}, but for the disk model with dark cloud dust.
\label{dist_ism_2}}
\end{figure}

\begin{figure}
\includegraphics[scale=0.7, angle=90]{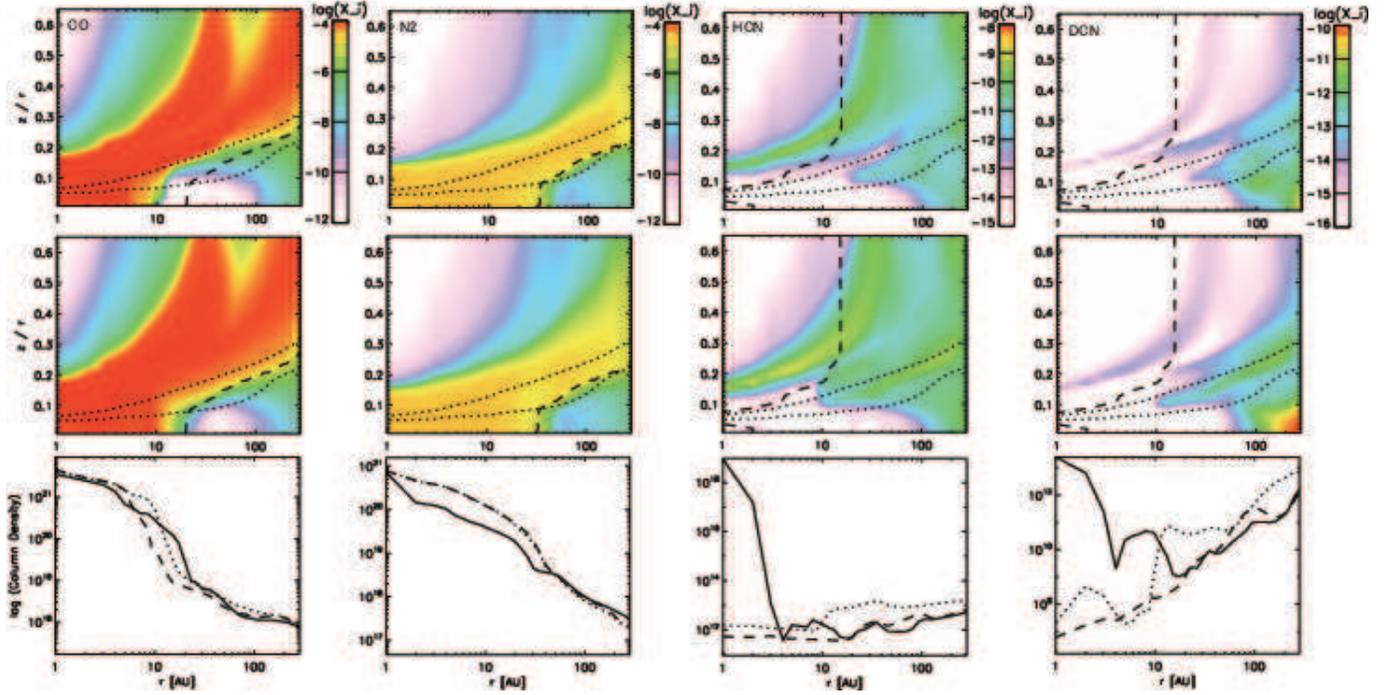}
\caption{Distributions of gaseous CO, N$_2$, HCN and DCN in the mm-sized grain model with vertical mixing at $t=3\times 10^5$ yr. The diffusion coefficient is
$\alpha=10^{-3}$ in the top panels, and $\alpha=10^{-2}$ in the middle panels. The bottom panels show the radial distribution of molecular column
densities of models without diffusion (solid) and with a diffusion coefficient of $10^{-3}$ (dashed), and $10^{-2}$ (dotted) at $t=3\times 10^5$ yr.
\label{diff_neutral}}
\end{figure}

\begin{figure}
\includegraphics[scale=0.7, angle=90]{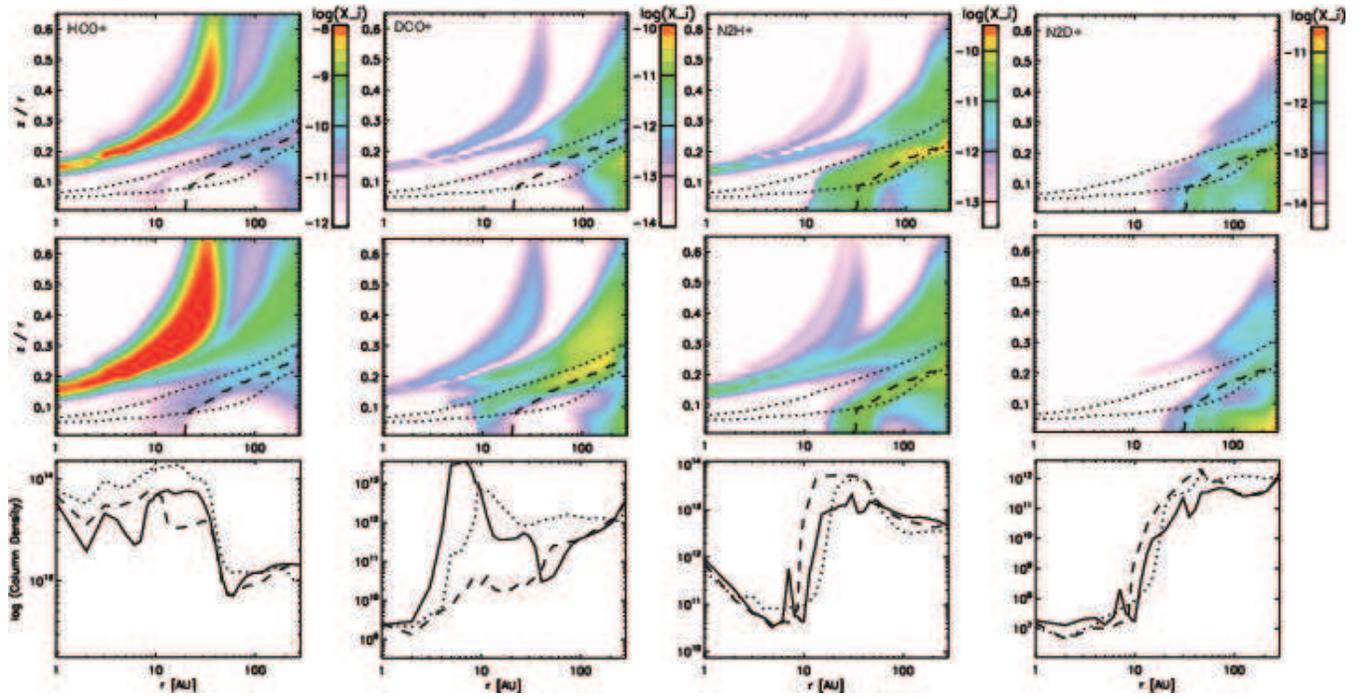}
\caption{Same as Figure \ref{diff_neutral}, but for ionic molecules.
\label{diff_ion}}
\end{figure}

\begin{figure}
\includegraphics[scale=0.7, angle=0]{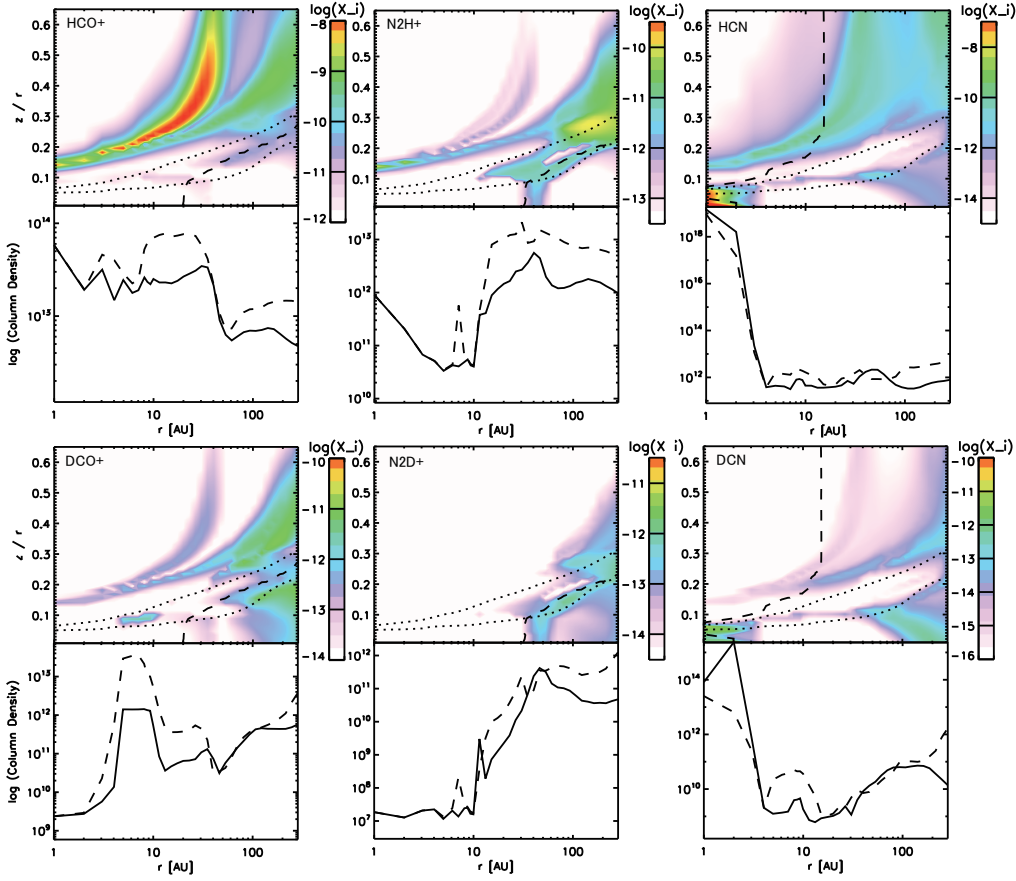}
\caption{Distributions of abundances and column densities of HCO$^+$, N$_2$H$^+$, HCN and their deuterated isotopologs in the mm-sized grain model without
cosmic-ray ionization at $t=3\times 10^5$ yr.
The solid lines depict the molecular column density without cosmic-ray ionization, while the dashed lines depict the column density in our fiducial model. 
\label{cr18}}
\end{figure}

Figure \ref{cr18} shows the distributions and column densities of HCO$^+$, N$_2$H$^+$, HCN and their deuterated isotopologs at $t=3\times 10^5$ yr. The solid lines depict the column density in the model without cosmic-ray ionization,
while the dashed lines depict our fiducial model. It is natural that the abundances of ionic molecules are suppressed by the low ionization rate.  
It should be noted that the dependence of
molecular ion column densities on cosmic-ray ionization rate would vary among disk models. In the model with dark cloud dust, the midplane abundances of these molecular ions are lower due to the more efficient freeze-out
of C-bearing and N-bearing molecules, and thus the X-ray dominated layer contributes more to the column density than in the model with mm-sized grains. Thus the decline of molecular ion column densities due to the exclusion 
of cosmic rays would be less significant in the disks with dark cloud dust.
We found that the column density of HCO$^+$ in the model without cosmic-ray ionization, for example,  is similar to that in Figure \ref{dist_ism_1} in the dark cloud dust model. The column density of DCO$^+$, on the other hand,
is reduced by about an order of magnitude at 10 au $\lesssim r \lesssim $ 100 au compared with Figure \ref{dist_ism_1}.

Comparing the right column in Figure \ref{cr18} with that in Figure \ref{dist_1mm_3}, we can see that both HCN and DCN are reduced in the outer ($r\gtrsim 10$ au) midplane
in the model without cosmic-ray ionization. Cosmic rays play a key role in forming N atoms and hydrocarbons such as CH$_3$ from N$_2$ and CO, respectively. N atoms are formed by the
recombination of N$_2$H$^+$ with a small branching ratio, which is produced by protonation of N$_2$. CO reacts with He$^+$ to form C$^+$, which goes through successive reactions with H$_2$ and electrons to form hydrocarbons.
Inside a radius of $\sim 3$ au, on the other hand, column densities of HCN and DCN are slightly higher in the model without cosmic rays than in Figure \ref{dist_1mm_3}, since the production of O atoms,
which destroy them, from water
and CO$_2$ is suppressed by the low ionization rate. 
In the disk with
dark cloud dust, both HCN and DCN are mostly abundant in the warm X-ray dominated layers, and thus their column densities do not depend much on the midplane ionization rate, except at $r\lesssim 3$ au,
where DCN is more abundant in the model without cosmic rays.

\section{Discussion}

The motivation of the present work is to investigate the major deuteration paths and their efficiency in disk models with an updated gas-grain chemical network, rather than constructing a best fit model for 
a specific object. It is, however, worth comparing our model results with  recent observations of deuterated species and their normal isotopologs. 
Table \ref{obs} summarizes the morphologies of integrated intensity maps of H$^{13}$CO$^+$, DCO$^+$, N$_2$H$^+$, N$_2$D$^+$, H$^{13}$CN, and DCN in two full disks around T Tauri stars (AS 209 and IM Lup),
two transition disks around T Tauri stars (V4046 and LkCa 15) and two disks around Herbig Ae stars (MWC 480 and HD 163296). It is mainly based on \cite{huang17}, and is supplemented by \cite{oberg11},
\cite{huang15}, \cite{salinas17}, and \cite{flaherty17}.
Comparison should be qualitative, rather than quantitative, since disk physical structures are expected to vary among objects, including the radial distribution of midplane temperature, which sets the location
of the CO snow line. While we compare the distribution of molecular lines with the estimated position of the CO snow line and temperature distribution for some disks, 
the derivation of temperature distributions in disks is not straightforward due to the opacity effect and temperature gradient in the vertical direction.
CO sublimation temperature, and thus the position of CO snow line could also depend on the ice composition, i.e. whether its surface is water-rich or not, while desorption energy of a molecule is set to be constant in our model.
It should also be noted that we compare the radial profiles (distributions) of molecular column density in our models with the observed distributions of line emissions.
\cite{huang17} observed the $J=3-2$ emission lines of DCO$^+$, DCN, H$^{13}$CO$^+$, and H$^{13}$CN, while \cite{huang15} observed the $J=3-2$ line of N$_2$D$^+$.
Using the RADEX code \citep{radex07}, we estimate that the opacity of the $J=3-2$ emission lines of HCO$^+$, DCO$^+$, N$_2$H$^+$ and HCN reaches unity, when the molecular column density is
a few $10^{12}$ cm$^{-2}$, with a gas temperature of 30 K. These emission lines from our model disks are thus expected to be optically thin except for limited
regions where the molecular column densities have a peak; the radial profile of the emission (e.g. ring emission) basically reflects the column density distributions.

\subsection{Molecular Ions}

In  IM Lup, DCO$^+$ emission shows a double-ring structure with radii of 110 au and 310 au, while H$^{13}$CO$^+$ has a single ring structure located at $\sim$ 130 au.
\cite{oberg15} attributed the H$^{13}$CO$^+$ ring and the inner ring of DCO$^+$ to the CO snow line. The central dip (i.e. a small region of low brightness)
in the molecular emission lines is suggested to be caused by
the subtraction of the optically thick dust continuum. Thus we cannot tell if H$^{13}$CO$^+$ and DCO$^+$ decrease inwards at $r\lesssim 95$ au. 
The location of the DCO$^+$ outer ring coincides with the edge of the disk observed in the millimeter continuum. \cite{oberg15} argued that in this outermost radius the dust opacity is reduced, which enhances
the photodesorption of CO and thus the DCO$^+$ abundance. Our model with mm-sized grains is consistent with this observation. The HCO$^+$ and DCO$^+$ column density increases inwards around (slightly outside)
the CO snow line. Beyond the CO snow line, on the other hand, the DCO$^+$ column density increases outwards, as in the outer ring in IM Lup, although desorption via cosmic-rays
is more efficient than photodesorption in the cold midplane in our model.

Another T Tauri disk, AS 209, shows more compact  emission of H$^{13}$CO$^+$ and DCO$^+$ than does IM Lup. While H$^{13}$CO$^+$ shows a ring of $r\sim 50$ au, the radial size of the DCO$^+$ emission is similar to or slightly extended
than 90 au, at which continuum emission shows a break. DCO$^+$ also has a central dip, which is shallower than that of H$^{13}$CO$^+$ \citep{huang17}. The midplane dust temperature distribution is estimated
from continuum observations \citep{andrews09}; it is about 20 K at $r\sim 60$ au. Hence the ring-like emission of H$^{13}$CO$^+$ could coincide with the CO snow line. Our models indicate that DCO$^+$ inside the
CO snow line could be formed via reactions (\ref{hcopd}) and CH$_4$D$^+$ + CO.
\cite{perez12} found that the dust opacity index $\beta$ increases outwards, which suggests that grains have grown (at least) to mm-sizes inside 80 au, while grains are still small at the outer radius.
The absence of an outer DCO$^+$ ring could be due to efficient freeze out (and chemical conversion) of CO molecules on small dust grains. C$^{18}$O emission, however, shows a local peak
at $r\sim 150$ au, indicating enhanced non-thermal desorption of CO there \citep{huang16}. It is not straightforward, either, to explain in our model why the central dip of H$^{13}$CO$^+$ is more clear than that of DCO$^+$.
Although the HCO$^+$ abundance in the midplane could decline at small radii due to a low ionization degree, or sublimation of H$_2$CO, H$_2$S and H$_2$O, which have higher proton affinities than CO, HCO$^+$ is the dominant charge carrier at the disk 
surface, which makes the distribution of  HCO$^+$ column density relatively flat.
In our model with CO and H$_2$O depletion, the HCO$^+$ column density shows a ring-like structure with a depression at $r\lesssim$ a few au, which indicates that the disk surface might be deficient in carbon at the dip radius.
A disk structure model specific for AS 209 and a chemical model with isotope selective CO photodissociation might also be necessary
to account for the central dip of H$^{13}$CO$^+$.

In addition to HCO$^+$, N$_2$D$^+$ is detected in AS 209. Its emission is offset from the disk center, and has a peak around the outer edge of the $^{13}$CO emission, which is consistent with our models.

LkCa15 is a transient disk with a central hole of $40-50$ au in the dust continuum. Distributions of H$^{13}$CO$^+$ and DCO$^+$ emission in LkCa15 are qualitatively similar to those in IM Lup;
the H$^{13}$CO$^+$ emission is rather compact, peaking at $\sim 40$ au, while the DCO$^+$ emission is more extended than the dust disk of radius $\sim 200$ AU, although the ring-like structure is much less clear in LkCa 15.
\cite{pietu06} derived a disk temperature of 22$\pm1$ K at $r=100$ au. Thus DCO$^+$ emission comes from both inside and outside the CO snow line. The ring of H$^{13}$CO$^+$ emission, on the other hand, could be linked to
the hole seen in the dust continuum.

Another transient disk, V4046, has a central hole of radius $\sim 29$ au in the dust continuum. While H$^{13}$CO$^+$ emission is diffuse extending to $\sim 200$ au, DCO$^+$ emission has a ring-like structure
with its peak at $\sim 70$ au \citep{huang17}. \cite{rosenfeld12} estimated the temperature distribution to be $T(r)=115 (r/10 {\rm au})^{-0.63}$ K; i.e. the temperature is 27 K at $r=100$ au. The DCO$^+$ ring thus could be
caused by  reactions (\ref{hcopd}) and CH$_4$D$^+$ + CO. It is consistent with our model that the H$^{13}$CO$^+$ distribution is rather flat and extends inwards compared with that of DCO$^+$. \cite{rosenfeld13}
determined that most of the dust mass is confined to a ring with a peak at $r=37$ au and FWHM of 16 au. The disk may not extend beyond the CO snowline, where DCO$^+$ could have another (outer)
emission peak.

In MWC 480, both H$^{13}$CO$^+$ and DCO$^+$ emissions have a peak at $\sim 40$ au \citep{huang17}. Despite the relatively high luminosity of the central star (11.5 $L_{\odot}$), the disk temperature is rather low.
\cite{pietu06} constrained the disk temperature to be $\sim 20$ K at a radius of $20-30$ au.  \cite{akiyama13} observed $^{12}$CO ($J=3-2$),  $^{12}$CO ($J=1-0$) and  $^{13}$CO ($J=1-0$), and estimated a gas temperature
of $\sim 13-15$ K for the layer traced by $^{13}$CO ($J=1-0$) at the radius of 100 au.
HCO$^+$ and DCO$^+$ emission thus seems to originate in the region around and slightly outside the CO snow line. 

In HD 163296, the H$^{13}$CO$^+$ ($J=3-2$) line features an emission ring peaking at $r \sim 50$ au, and also has a more extended component with a break at $r\sim 200$ au.
The DCO$^+$ emission shows a rather broad ring with a peak at $r\sim 70$ au \citep{huang17}. The temperature distribution in the disk of HD 163296 was derived by \cite{isella16} from CO observations;
the midplane temperature is 23 K at $r \sim 80$ au. \cite{qi15}, on the other hand, derived the location of the CO snow line to be $r\sim 90$ au from observations of N$_2$H$^+$ and CO isotopologs.
HCO$^+$ and DCO$^+$ emission thus originates both inside and outside of the CO snow line. 
Recently, \cite{salinas17} reported spatially resolved emissions of DCO$^+$, N$_2$D$^+$, and DCN. Their analysis shows that DCO$^+$ emission can be divided to three ring-like components at 70 au, 150 au, and 260 au \citep[see also][]{flaherty17}.
The innermost and sencond ring coincide with the emission peaks of DCN and N$_2$D$^+$, respectively, while the outermost ring could arise from the non-thermal desorption of CO, as in the outer ring of IM Lup
\citep{salinas17}. In our fiducial model, the radial distribution of the DCO$^+$ column density can be divided to three components at $r\sim$ several au, 10-30 au, and an outermost radius ($\gtrsim 100$ au).
The innermost peak coincides with the local peak of DCN, while DCO$^+$ originates in non-thermal desorption of CO in the midplane at the outermost radius, which could be consistent with \cite{salinas17}.
The second component at $r\sim 10-30$ au, however, is located inside the local peak of the N$_2$D$^+$ column density ($\gtrsim 20-100$ au).

So far, the DCO$^+$ emission detected in disks always has a central hole or dip \citep{huang17, qi15}.
In our models, the DCO$^+$ distribution does have a dip.  Right outside the dip, DCO$^+$ is mainly formed by  reaction (\ref{hcopd}), while D atoms are destroyed mainly by the reaction with HS inside the dip.
Although the sublimation of S-bearing species such as H$_2$S seems to be a key for the decline of D atom in our model, a derivation of the physical condition for the decline of D atoms and thus of DCO$^+$
may not be straightforward, since D atoms are chemically active. It should also be noted that a central dip could be caused by subtraction of
optically thick dust continuum, rather than by a chemical effect, in some disks \citep{huang17, salinas17}.



\subsection{HCN and DCN}

In our models, both in models with mm-sized dust and dark cloud dust, the radial distributions of the HCN and DCN column densities are centrally peaked, although the
DCN/HCN ratio is low ($10^{-5}-10^{-4}$) at the central region ($r\lesssim 10$ au) due to high temperature. Their column densities drop sharply outside this high temperature region, and then slightly
increase outwards, with the gradient steeper for DCN. In the models with vertical diffusion, the central peak disappears, since hydrocarbons, from which HCN and DCN are formed, are transported to the disk surface to be destroyed.

In the disk of IM Lup, H$^{13}$CN is not detected and DCN emission is weak and diffuse \citep{huang17}, 
while both H$^{13}$CN and DCN are clearly detected and have a central dip \citep{huang17} in the disk of AS 209.
In these disks, HCN and DCN are not centrally peaked, possibly because the temperature is not high enough to desorb HCN at the radius traced by current observations
(beam size of $\sim 60$ au), or because turbulent mixing is at work.
V4046 and HD163296, on the other hand, show  centrally peaked H$^{13}$CN emission and a central dip in DCN emission.
If these features simply reflect their column density distributions, it is difficult to account for in our fiducial model, in which both HCN and DCN are most abundant in the innermost radii.
Our models with turbulent diffusion, however, might produce the observed feature; while DCN is abundant only in the outer cold regions, HCN is abundant
at the disk surface even inside 20 au, which could be bright due to high temperatures.

\section{SUMMARY}

We investigated deuterium chemistry coupled with the nuclear spin-state chemistry of H$_2$ and H$_3^+$ and their isotopologs in protoplanetary disks. Our principal
findings are as follows:
\begin{enumerate}
\item{We have found multiple paths for deuterium enrichment.  The exchange reactions with D atoms, such as HCO$^+$ + D, are found to be effective,
in addition to  H$_3^+$ + HD, CH$_3^+$ + HD, and C$_2$H$_2^+$ + HD, which had been considered in previous studies.}

\item{As discussed in Appendix A, the OPR of H$_2$ is found to be almost thermal, as long as the cosmic rays ionize the disk with a rate $\sim 10^{-17}$ s$^{-1}$.
In the cold ($\sim$ 10 K) midplane, however, the OPR reaches its minimum value,
which is higher than the thermal value. The minimum value is determined by the balance between the rates of H$_2$ formation, which sets the OPR to be 3,
and spin conversion via ion-molecule reactions involving protons and H$_3^+$ mainly.
The OPR could reach the thermal value at such cold dense regions, if we take into account the spin conversion on grain surfaces, which has recently been found in
laboratory experiments \citep{ueta16}. We have also analyzed the OPR of H$_3^+$ and H$_2$D$^+$,
and the abundances of H$_2$D$^+$, DCO$^+$, and N$_2$D$^+$ in the cold midplane, in part by the use of derived  analytical formulae.}

\item{In our models, the contribution of the exchange reaction CH$_3^+$ + HD is found to be less significant than that described in \cite{favre15}. The increasing OPR of H$_2$
helps the backward reaction at $T\gtrsim 20$ K.}

\item{In the disk model with mm-sized grains, reduced freeze-out rates enhance the gaseous molecular abundances in the cold midplane. 
DCO$^+$, N$_2$D$^+$, and DCN in the outer midplane thus contribute significantly to their column densities. The radial distribution
of the DCO$^+$ column density has a double-ring structure, similar to the DCO$^+$ emission observed in IM Lup. While the outer ring is caused by the enhanced deuteration
of H$_3^+$ and less efficient chemical conversion of HD and CO in the outer radii in our model,
the inner ring is linked to the CO snow line and depletion of D atoms due to reactions with HS for example. N$_2$D$^+$, on the other hand, is more
abundant outside the CO snow line. Gaseous DCN decreases inward, except in the central hot region ($T\gtrsim 100$~K) where it is thermally desorbed.}

\item{If the elemental C/O ratio is higher than unity due to sedimentation of H$_2$O ice, hydrocarbons become abundant. The exchange reaction CH$_3^+$ + HD,
which eventually leads to the ion CH$_4$D$^+$,  thus contributes to
form DCO$^+$ via the reaction CH$_4$D$^+$ +CO in the warm molecular layer. The column densities of gaseous HCN and DCN are also enhanced by an order of magnitude compared with
the fiducial model. The spatial distributions of molecular abundances and radial profiles of molecular column densities are, however, qualitatively similar to those in our fiducial model.
}

\item{In the disk model with dark cloud dust, the freeze out of molecules is more severe in the outer midplane, while the disk surface is better
shielded from UV radiation than in the model with mm-sized grains. The disk surface area thus harbors abundant gaseous molecules and contributes
to the column densities (and emissions) of HCO$^+$, DCO$^+$, N$_2$H$^+$, HCN and DCN.
One exception is N$_2$D$^+$, which is not abundant in the disk surface.}

\item{Turbulence helps to prevent chemical conversion of molecules to less volatile species by transporting ices from the midplane to the
disk surface, where ices are desorbed and photodissociated, but it also enhances the formation of saturated or less volatile molecules in the midplane by transporting
H atoms and radicals from the disk surface. For example, NH$_3$ ice formation is hampered and thus N$_2$ and N$_2$H$^+$ abundances tend to be
enhanced inside the N$_2$ snow line. Turbulence also transports D atoms in the disk surface
to the lower layers, which helps the formation of DCO$^+$ and N$_2$D$^+$. At the innermost radii, the abundances of HCN and DCN are
significantly reduced by the turbulence; their icy components and hydrocarbons are transported to the disk surface and destroyed, while O atoms are
transported to the midplane to react with C-bearing species.}

\item{If the penetration of cosmic rays is  hampered by the stellar wind, the midplane ionization rate decreases by more than one order of magnitude.
Column densities of molecular ions decline, although the decrement varies with radius, species, and the dust grain sizes in the disk model.
HCN and DCN also decrease at $r >$ several au,
since the cosmic-ray ionization is needed to form their precursors, N atoms and hydrocarbons, from N$_2$ and CO, respectively.}

\end{enumerate}

\acknowledgments

This work is supported by JSPS KAKENHI Grant Numbers 23540266, 16H00931, and 17K14245. We would like to thank Hideko Nomura for providing the disk physical models, and to Jane Huang, Megumi Kayanuma, and Liton Majumdar for helpful discussions.
We are grateful to the anonymous referee for helpful comments, which improved the manuscript.
E. H. wishes to acknowledge the support of the National Science Foundation through grant AST-1514844.

\appendix
\section{H$_2$ OPR}

In our chemical reaction network, the OPR is assumed to be 3 for newly formed H$_2$ molecules on grain surfaces. Neglecting rapid thermalization on the grain, they are
desorbed into the gas phase, where the spin state changes via proton exchange reactions with ions such as H$_3^+$.
Figure \ref{H2opr} (a) shows the 2D distribution of gas temperature, and Figure \ref{H2opr} (b) shows the 2D distribution of the H$_2$ OPR
at $t=3\times 10^5$ yr in our fiducial model. The distribution is the same at $t=1\times 10^5$ yr, $3\times 10^5$ yr and $9.3 \times 10^5$,
because the H$_2$ OPR reaches steady state on a relatively short timescale ($< 10^5$ yr).
In Figure \ref{H2opr} (a) and (b) the horizontal axis ($r$) is shown in linear scale, since we discuss only the low temperature
regions in this Appendix. The OPR apparently follows the gas temperature distribution.
In order to study the dependence of OPR on gas temperature more quantitatively, we plot the OPR as a function of 
gas temperature in Figure \ref{H2opr} (c). At low temperatures ($\lesssim$ several tens of K), where only the lowest two rotational levels need be considered,
the OPR is mostly equal to the value of thermal equilibrium,
\begin{equation}
n({\rm o\mathchar`- H_2})/n({\rm p\mathchar`-H_2})=9\exp (-170.5/T), \label{a_oprH2}
\end{equation}
which is depicted by the solid line, while at higher temperatures the OPR asymptotically reaches the statistical equilibrium value of 3.
At $T\sim 10$ K, however, the OPR obtained in the numerical calculation reaches minimum value, which is higher than the thermal value. The minimum value can be explained as follows.

The steady state value of the H$_2$ OPR is given by
\begin{eqnarray}
{\rm OPR}_{\rm st}&=&\frac{\beta_1+b_0 \beta_2}{1+(1-b_0)\beta_2} \label{H2opr_beta}\\
\beta_1&=&\tau_{\rm op}/\tau_{\rm po} \nonumber \\
\beta_2&=&\tau_{\rm op}/\tau_{\rm H2} \nonumber,
\end{eqnarray}
where $b_0$ ($=0.75$) is the branching ratio to form o$\mathchar`-$H$_2$ for H$_2$ formation on grains,
$\tau_{\rm H2}$ is the timescale of H$_2$ formation, and $\tau_{\rm op}$ and $\tau_{\rm po}$ represent the timescale of
spin conversion via proton exchange from ortho to para and from para to ortho, respectively \citep[e.g.][]{furuya15}. When the abundance of ions such as H$_3^+$
is high enough, $\beta_1$ dominates in (\ref{H2opr_beta}), and the OPR reaches the thermal value. If the ion abundance is low, on the other hand,
the OPR is determined by $\beta_2$; i.e. the balance between the H$_2$ formation, which sets the OPR to 3, and the spin conversion
via ion-molecule reactions. In the disk model, the low temperature ($T\sim 10$ K) region corresponds to the outer midplane; due to the low ionization degree, the OPR
is set by $\beta_2$:
\begin{equation}
\beta_2=\frac{\pi a^2 \sqrt{\frac{2kT}{\pi m}}n ({\rm grain}) n({\rm H})}{k_{\rm op}n({\rm XH}^+)n({\rm H_2})} \nonumber,
\end{equation}
where $k$ is the Boltzmann constant, $m$ is the mass of a hydrogen atom, $n$(XH$^+$) is the number density of ions which convert ortho- to
para-H$_2$ through proton exchange reactions with a rate coefficient $k_{\rm op}$. Considering the balance between the formation and destruction of H$_2$,
the numerator, i.e. the formation rate of H$_2$, can be replaced by the H$_2$ destruction rate via cosmic-ray/X-ray ionization;  
\begin{equation}
\beta_2=\frac{\zeta n({\rm H}_2)}{k_{\rm op}n({\rm XH}^+)n({\rm H_2})}=\frac{\zeta}{k_{\rm op}n({\rm XH}^+)}, \label{beta2}
\end{equation}
where $\zeta$ is the H$_2$ ionization rate. Then the minimum value is almost equivalent to the OPR given by \cite{walmsley04}.
It should also be noted that the minimum value of OPR in our disk model is much lower than $\sim 10^{-3}$, which is often assumed in the model of molecular clouds \citep[e.g.][]{flower06, faure13},
because $n$(XH$^+$) in eq. (\ref{beta2}) depends on the ionization rate and gas density.

Recent laboratory experiments found that the OPR of H$_2$ is thermalized via the spin conversion on amorphous water ice \citep{ueta16}. The conversion rate is higher than
the desorption rate at $\lesssim 12$ K. Thus OPR could reach the thermal value even at 10 K, if we include the conversion on grain surfaces.

Figure \ref{H2opr} (c) shows that there are some deviations of the H$_2$ OPR from the thermal value at warm temperatures ($T\gtrsim 25$ K), as well.
Such deviations are caused by the relatively low ionization degree in the warm dense midplane. The OPR cannot simply be calculated from $\beta_2$ (eq. \ref{beta2}) in these regions, since
$\beta_1$ and $\beta_2$ are comparable, unlike the low temperature limit ($T\sim 10$ K) discussed above.

Finally it should be noted that, in the model without cosmic-ray ionization, the region with non-thermal H$_2$ OPR extends both in the radial and vertical directions (Figure \ref{H2opr} (d)).
The ionization rate is too low around the midplane for $\beta_1$ to dominate in equation (\ref{H2opr_beta}).

\section{Analytical formula of the OPR values and abundances of molecular ions in the outer midplane}

In \S 3, we showed that the major formation paths of deuterated molecular ions vary spatially. It is thus difficult to derive analytical
formulae of abundances of these ions which are applicable to the whole disk. Derivation of the analytical formulas is possible, however, if we
restrict ourselves to the cold outer midplane. Since molecules such as H$_2$D$^+$ and N$_2$D$^+$ have their peak abundances in the outer midplane,
and since they are often considered to be  tracers of ionization degree, the analytical formulae could be useful.

\begin{figure}
\includegraphics[scale=0.8, angle=0]{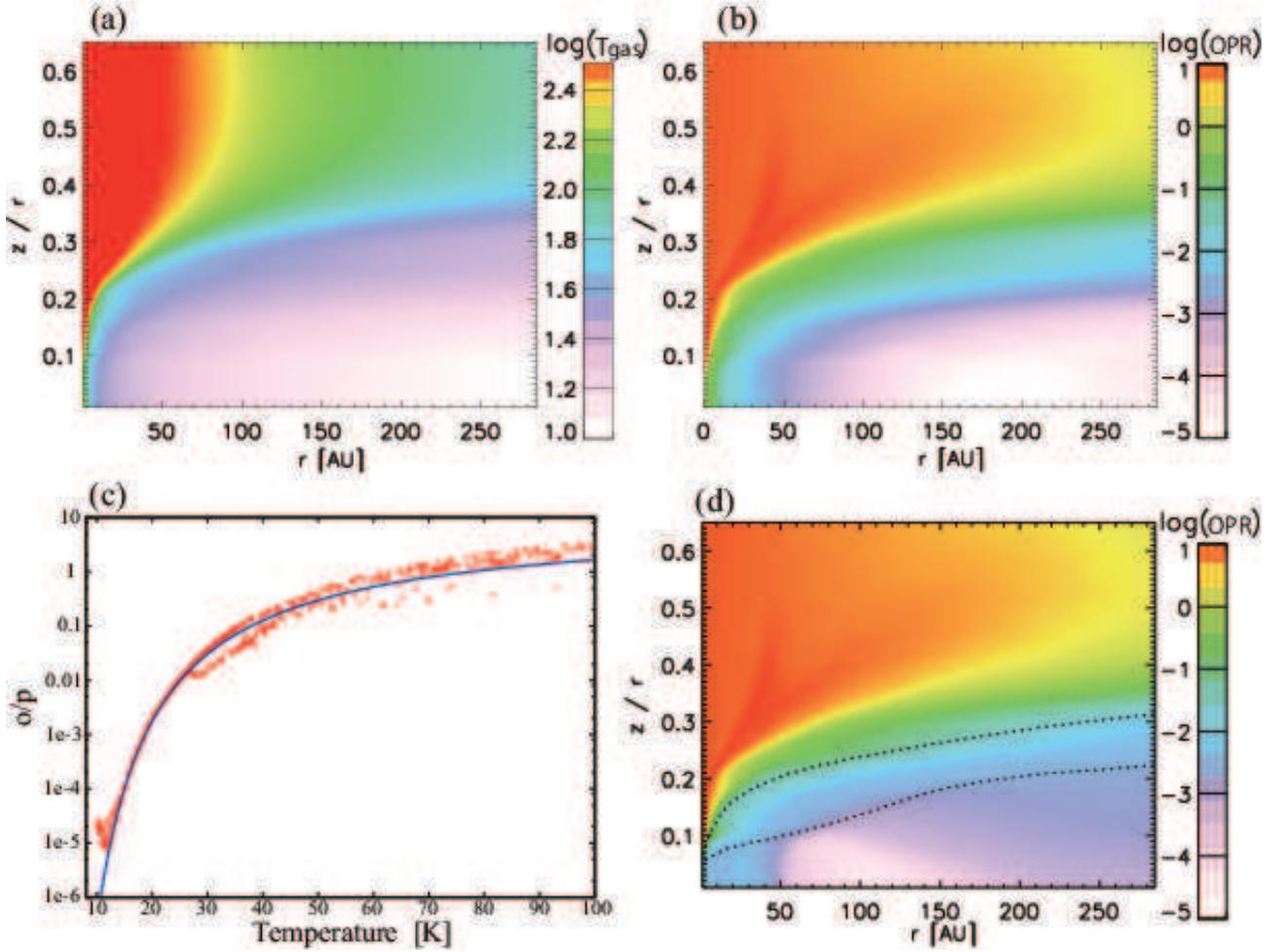}
\caption{(a) Distribution of gas temperature in our fiducial disk model. (b) Distribution of H$_2$ OPR in our fiducial disk model. (c) H$_2$ OPR in our fiducial model (red crosses) and thermal equilibrium value (blue) as functions of temperature.
(d) Same as (b), but for the model without cosmic-ray ionization. The upper and lower dotted lines depict the positions where the X-ray ionization rate is equal to $5\times 10^{-17}$ s$^{-1}$ and $1\times 10^{-18}$ s$^{-1}$, respectively.
\label{H2opr}}
\end{figure}

\begin{figure}
\includegraphics[scale=0.8, angle=0]{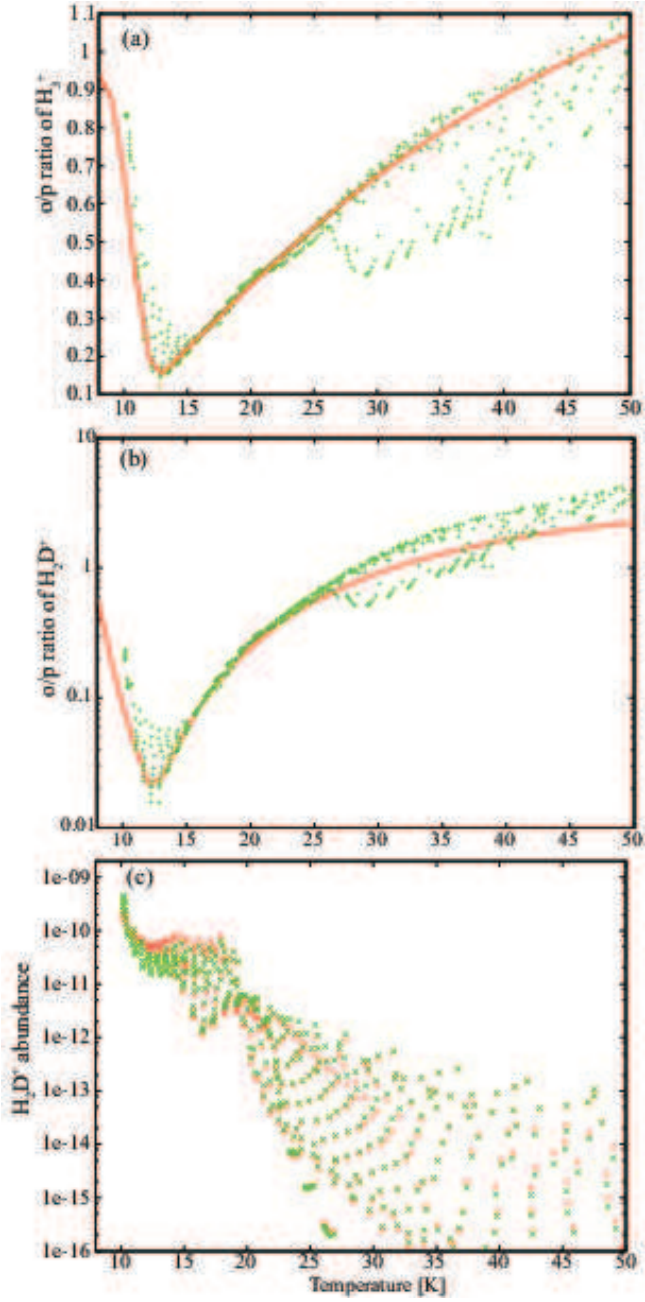}
\caption{(a) OPR of H$_3^+$ in our fiducial model (green crosses) and thermal equilibrium value (red) as a function of temperature.
(b) The o/p ratio of H$_2$D$^+$ in our fiducial model (green) and the analytical value (red).
(c) H$_2$D$^+$ abundance in our fiducial model (green) and the value obtained by the analytical formula (red).
\label{opr_H3p}}
\end{figure}

First, we derive the OPR of H$_3^+$. The analytical formula for the H$_3^+$ abundance is given in eq (18) in \cite{aikawa15}; 
it is determined by the formation via cosmic-ray/X-ray ionization of H$_2$ and the destruction by reactions with CO, N$_2$, electrons, and negatively charged grains. Its spin state changes via reactions with H$_2$
\citep{oka04}:
\begin{eqnarray}
{\rm o\mathchar`-H_3^+ + p\mathchar`-H_2 \rightarrow p\mathchar`-H_3^+ + o\mathchar`-H_2} \\
{\rm o\mathchar`-H_3^+ + o\mathchar`-H_2 \rightarrow p\mathchar`-H_3^+ + o\mathchar`-H_2} \\
{\rm p\mathchar`-H_3^+ + o\mathchar`-H_2 \rightarrow o\mathchar`-H_3^+ + p\mathchar`-H_2} \\
{\rm p\mathchar`-H_3^+ + o\mathchar`-H_2 \rightarrow o\mathchar`-H_3^+ + o\mathchar`-H_2}.
\end{eqnarray}
The OPR of H$_3^+$ is thus given at steady state by
\begin{equation}
\frac{n{\rm (o\mathchar`-H_3^+)}}{n{\rm (p\mathchar`-H_3^+)}}=\frac{(k_{\rm B3}+k_{\rm B4})n{\rm (o\mathchar`-H_2)}}{k_{\rm B1}n{\rm (p\mathchar`-H_2)}+k_{\rm B2}n{\rm (o\mathchar`-H_2)}}. \label{a_oprH3p}
\end{equation}
Figure \ref{opr_H3p} (a) shows the H$_3^+$ OPR in our fiducial disk model (green crosses) and the value given by equation (\ref{a_oprH3p}) (red line). We fix the H$_2$ OPR
of (\ref{a_oprH2}) and its minimum value of $1\times 10^{-5}$, for simplicity. We see reasonable agreement between (\ref{a_oprH3p}) and the OPR of H$_3^+$ in the numerical calculation,
while there are some deviations at $T\gtrsim 25$ K, which originate from the non-thermal OPR of H$_2$ (Figure \ref{H2opr} c).
The deviation stands out in Figure \ref{opr_H3p} (a), in which the vertical axis is shown in linear scale, while it is shown in logarithmic scale in Figure \ref{H2opr} (c).

The major formation and destruction paths of H$_2$D$^+$ are
\begin{eqnarray}
{\rm o\mathchar`-H_3^+ + HD \rightarrow o\mathchar`-H_2D^+ + o\mathchar`-H_2} \label{a1}\\
{\rm p\mathchar`-H_3^+ + HD \rightarrow o\mathchar`-H_2D^+ + p\mathchar`-H_2} \label{a2}\\ 
{\rm p\mathchar`-H_3^+ + HD \rightarrow p\mathchar`-H_2D^+ + o\mathchar`-H_2} \label{a4}\\
{\rm p\mathchar`-H_3^+ + HD \rightarrow p\mathchar`-H_2D^+ + p\mathchar`-H_2} \label{a5}\\ 
{\rm o\mathchar`-H_2D^+ + o\mathchar`-H_2 \rightarrow o\mathchar`-H_3^+ + HD} \label{a3}\\
{\rm p\mathchar`-H_2D^+ + o\mathchar`-H_2 \rightarrow o\mathchar`-H_3^+ + HD} \label{a6}\\
{\rm p\mathchar`-H_2D^+ + o\mathchar`-H_2 \rightarrow p\mathchar`-H_3^+ + HD} \label{a7}\\
{\rm o\mathchar`-H_2D^+ + CO \rightarrow HCO^+ + HD}\label{a8}\\
{\rm p\mathchar`-H_2D^+ + CO \rightarrow HCO^+ + HD}\label{a9}\\
{\rm o\mathchar`-H_2D^+ + CO \rightarrow DCO^+ + o\mathchar`-H_2}\label{a10}\\
{\rm p\mathchar`-H_2D^+ + CO \rightarrow DCO^+ + p\mathchar`-H_2}. \label{a11}
\end{eqnarray}
The balance between these reactions is described as
\begin{eqnarray}
n({\rm HD})\{k_{\ref{a1}}n({\rm o\mathchar`-H_3^+})+(k_{\ref{a2}}+k_{\ref{a4}}+k_{\ref{a5}})n({\rm p\mathchar`-H_3^+})\}&=&\\ \nonumber
n({\rm o\mathchar`-H_2}) \{k_{\ref{a3}}n({\rm o\mathchar`-H_2D^+}) &+& (k_{\ref{a6}}+k_{\ref{a7}})n({\rm p\mathchar`-H_2D^+})\} \\ \nonumber
 + (k_{\ref{a8}}+k_{\ref{a10}})n({\rm o\mathchar`-H_2D^+})n({\rm CO}) &+&  (k_{\ref{a9}}+k_{\ref{a11}})n({\rm p\mathchar`-H_2D^+})n({\rm CO}).
\end{eqnarray}
The spin state of H$_2$D$^+$ is converted via
\begin{eqnarray}
{\rm p\mathchar`-H_2D^+ + o\mathchar`-H_2 \rightleftharpoons o\mathchar`-H_2D^+ + p\mathchar`-H_2} \label{b1} \\
{\rm o\mathchar`-H_2D^+ + o\mathchar`-H_2 \rightleftharpoons p\mathchar`-H_2D^+ + p\mathchar`-H_2} \label{b2} \\
{\rm o\mathchar`-H_2D^+ + o\mathchar`-H_2 \rightleftharpoons p\mathchar`-H_2D^+ + o\mathchar`-H_2} \label{b3}
\end{eqnarray}
in a shorter timescale than that of H$_2$D$^+$/H$_3^+$ conversion. The o/p ratio of H$_2$D$^+$ can thus be described by
\begin{equation}
\frac{n({\rm o\mathchar`-H_2D^+})}{n({\rm p\mathchar`-H_2D^+})}=\frac{(k_{\ref{b1}}^++k_{\ref{b3}}^-) \times n({\rm o\mathchar`-H_2})/n({\rm p\mathchar`-H_2})+
k_{\ref{b2}}^-}{(k_{\ref{b2}}^++k_{\ref{b3}}^+) \times n({\rm o\mathchar`-H_2})/n({\rm p\mathchar`-H_2})+k_{\ref{b1}}^-} \label{opr_H2Dp}
\end{equation}
\citep{gerlich02, brunken14}. The
H$_2$D$^+$ abundance is then given by
\begin{equation}
n({\rm H_2D^+})=\frac{ n({\rm HD})\{k_{\ref{a1}}n({\rm o\mathchar`-H_3^+})+(k_{\ref{a2}}+k_{\ref{a4}}+k_{\ref{a5}})n({\rm p-H_3^+})\}}{k_{\ref{a3}}n({\rm o-H_2})\frac{x}{x+1}
+(k_{\ref{a6}}+k_{\ref{a7}})n({\rm o\mathchar`-H_2})\frac{1}{x+1}+(k_{\ref{a8}}+k_{\ref{a10}})n({\rm CO})}, \label{Ah2dp}
\end{equation}
where $x$ is the o/p ratio of H$_2$D$^+$ (\ref{opr_H2Dp}). Note that the rate coefficient of H$_2$D$^+$ + CO is the same for o$\mathchar`-$H$_2$D$^+$ and p$\mathchar`-$H$_2$D$^+$.
If the H$_2$D$^+$ abundance derived above is comparable to that of main isotopolog, H$_3^+$, it should be replaced by
\begin{equation}
n({\rm H_2D^+})=n({\rm H_3^+})\frac{n'({\rm H_2D^+})}{n({\rm H_3^+})+n'({\rm H_2D^+})},
\end{equation}
where $n'({\rm H_2D^+})$ denotes the value obtained in (\ref{Ah2dp}).

Reasonable agreement between the analytical value and the numerical results for the o/p ratio of H$_2$D$^+$ is found in Figure \ref{opr_H3p} (b).
In our numerical models, multiply-deuterated H$_3^+$ becomes more abundant than H$_2$D$^+$ in the outermost ($r\gtrsim 250$ au) midplane.
The analysis of multi-deuterated species is, however, beyond the scope of this paper.  Since we neglect the multiply deuterated  isotopologs of H$_3^+$,
H$_2$D$^+$ represents the total deuterated-H$_3^+$ in our analysis. In the following, we therefore compare the analytical value of the H$_2$D$^+$ abundance with the sum of the abundances of deuterated H$_3^+$
($n({\rm D_3^+})/n_{\rm H} + n({\rm D_2H^+})/n_{\rm H} + n({\rm H_2D}^+)/n_{\rm H}$) obtained in our fiducial model.
Figure \ref{opr_H3p} (c) shows the sum of deuterated H$_3^+$ abundance in our fiducial model (green), which is compared with the
analytical value (red). At warm regions ($T\gtrsim 20$ K), the abundances of H$_3^+$ and its deuterated isotopologs are not simple functions of temperature; they depend on abundances of other species
such as CO, N$_2$ and electrons \citep{aikawa15}. We use the latter abundances from our fiducial model in the calculation of the analytical formula of H$_2$D$^+$ abundance. Equation (\ref{Ah2dp}) also needs the abundance of HD as an input. While its canonical value is $1.5\times 10^{-5}$, HD abundance is decreased down to $\sim 10^{-7}$ via conversion to D-bearing ices such as NH$_2$D and
HDO in the cold midplane in our numerical model \citep[see also][]{sipila13, teague15}.
We thus use the HD abundance from our fiducial model, as well.

\begin{figure}
\includegraphics[scale=0.8, angle=90]{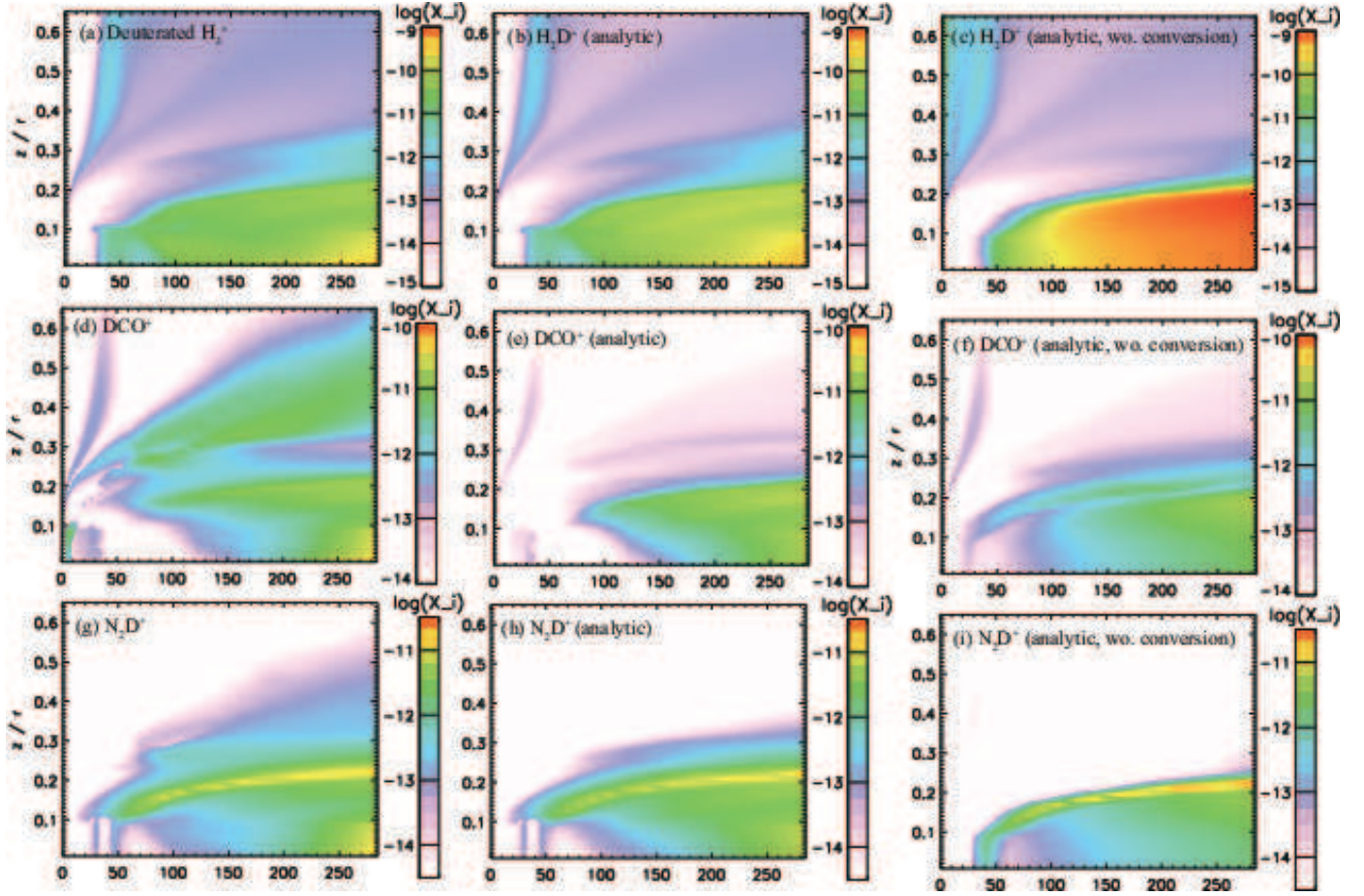}
\caption{Panels (a), (d), and (g) show distributions of deuterated H$_3^+$, DCO$^+$ and N$_2$D$^+$ in our fiducial model at $t=3\times 10^5$ yr.
Panels (b), (e), and (h) show distributions of H$_2$D$^+$, DCO$^+$ and N$_2$D$^+$ calculated by the analytical formulas
using the abundances of HD, CO, N$_2$, and electrons from the fiducial model. Panels (c), (f), and (i) show distributions of H$_2$D$^+$, DCO$^+$ and N$_2$D$^+$ calculated by the analytical formulas with the abundances of HD, CO and N$_2$
given analytically.
\label{a_H2Dp}
}
\end{figure}

Figures \ref{a_H2Dp} (a) and (b) show the 2D distribution of deuterated H$_3^+$ in our fiducial model, and the H$_2$D$^+$ abundance derived from eq. (\ref{Ah2dp}), respectively. We can see a reasonable agreement between
the numerical model and the analytical formula.
For a comparison, Figure \ref{a_H2Dp} (c) shows the distribution of H$_2$D$^+$ abundance derived from eq. (\ref{Ah2dp}) assuming that CO, N$_2$ and HD are not converted to other molecules; 
HD abundance is set to be constant ($n$(HD)/$n_{\rm H}=1.5\times 10^{-5}$), and CO and N$_2$ abundances are determined simply by the balance between adsorption and desorption \citep{aikawa15}.
H$_2$D$^+$ abundance is higher than the numerical model, mainly due to a higher HD abundance.


Once we obtain the analytical formula of the H$_2$D$^+$ abundance, we can calculate the DCO$^+$ and N$_2$D$^+$ abundances, by simply replacing H$_3^+$
with H$_2$D$^+$ and modifying the corresponding reaction rate coefficients in equations (22) and (21) in \cite{aikawa15}.
The resultant 2D distributions are shown in Figure \ref{a_H2Dp} (d-i). Panels (e) and (h) show distributions of DCO$^+$ and N$_2$D$^+$, respectively, derived from the analytical formulas using the abundances of HD, CO, N$_2$, and electrons from
our fiducial model, while the abundances of HD, CO and N$_2$ are given analytically for panels (f) and (i).
The agreements between the numerical model (panels d and g) and analytical values (panels e and h) confirm that DCO$^+$ and N$_2$D$^+$ are mainly formed via deuterated H$_3^+$ in the outer midplane.

\begin{table}
\begin{center}
\caption{Initial abundances of major species \label{initial}}
\begin{tabular}{ll}
\tableline\tableline
species & abundance\\
\tableline
N & 1.8 (-8) \tablenotemark{a} \\
O & 8.8 (-9)  \\
o-H$_2$ & 1.6 (-3)\\
p-H$_2$ & 5.0 (-1)\\
HD & 1.3 (-5) \\
CO & 6.2 (-5) \\
CH$_4$ & 7.4 (-6)\\
H$_2$CO & 3.5 (-6) \\
CO$_2$ & 2.6 (-6) \\
CH$_3$OH & 8.8 (-7) \\
N$_2$ & 3.8 (-6) \\
HCN & 2.0 (-7) \\
DCN & 7.7 (-9) \\
NH$_3$ & 1.7 (-5) \\
NH$_2$D & 2.0 (-7)\\
H$_2$O & 1.1 (-4) \\
HDO & 8.5 (-8) \\\tableline
\tablenotetext{a}{$a(b)$ means $a\times 10^{b}$.}
\end{tabular}

\end{center}
\end{table}

\begin{table}
\begin{center}
\begin{scriptsize}
\caption{Observations morphologies of molecular lines intensities\tablenotemark{a} \label{obs}}
\begin{tabular}{lccccccc}
\tableline\tableline
Object & Class & H$^{13}$CO$^+$ & DCO$^+$ & N$_2$H$^+$ & N$_2$D$^+$ & H$^{13}$CN & DCN \\
\tableline
IM Lup & full disk (T Tauri) & ring-like & multiple rings & offset\tablenotemark{b} & - & ring-like\tablenotemark{c} & extended?\tablenotemark{d}\\
AS 209 & full disk (T Tauri) & ring-like & ring-like & offset\tablenotemark{b} & offset\tablenotemark{e} & ring-like & ring-like \\
LkCa 15 & transition disk (T Tauri) & ring-like & diffuse & - & - & compact?\tablenotemark{d} & multiple rings?\tablenotemark{d} \\
V4046 Sgr & transition disk (T Tauri) & diffuse & ring-like & offset\tablenotemark{b} & -  & centrally peaked & ring-like \\
MWC 480 & Herbig Ae & ring-like & ring-like & - & - & centrally peaked & compact?\tablenotemark{d} \\
HD 163296 & Herbig Ae & multiple rings & multiple rings\tablenotemark{f,g}  & - & ring-like\tablenotemark{g}& centrally peaked & ring-like \\ \tableline
\tablenotetext{a}{Based on \cite{huang17} unless otherwise stated.}
\tablenotetext{b}{\cite{oberg11}}
\tablenotetext{c}{H$^{13}$CN is not detected. H$^{12}$CN shows a ring-like feature.}
\tablenotetext{d}{Detected but not categorized in \cite{huang17} due to relatively low S/N.}
\tablenotetext{e}{\cite{huang15}}
\tablenotetext{f}{\cite{flaherty17}}
\tablenotetext{g}{\cite{salinas17}}
\end{tabular}
\end{scriptsize}
\end{center}
\end{table}

\begin{table}
\begin{center}
\caption{Rate coefficients of major reactions of deuterated molecular ions\tablenotemark{a} \label{rates}}
\begin{tabular}{ll}
\tableline\tableline
reaction & rate coefficients [cm$^3$ s$^{-1}$]\\
\tableline
B1 & $1.5\times 10^{-9}\exp(-136/T)$\tablenotemark{b} \\
B2 & $4.00 \times 10^{-10}\exp(0.19/T)$\\
B3 & $3.46\times 10^{-10} \exp(0.69/T)$\\
B4 & $8.03\times 10^{-10}\exp(-32.6/T)$\\
B6 & $1.11\times 10^{-9}\exp(-0.35/T)$\\
B7 & $6.08\times 10^{-10}\exp(1.08/T) $\\
B8 & $4.93\times 10^{-10}\exp(0.95/T) $\\
B9 & $3.11\times 10^{-10}\exp(0.71/T) $\\
B10 & $1.64\times10^{-10}\exp(-6.31/T) $\\
B11 & $9.32\times 10^{-9}\exp(-94.6/T) $\\
B12 & $1.48\times 10^{-10}\exp(-58.8/T) $\\
B13 & $1.07\times 10^{-9} $\\
B15 & $5.37\times 10^{-10} $\\
B18$^+$ & $ 1.26\times 10^{-9}\exp(-0.06/T)$\\
B18$^-$ & $ 5.58\times 10^{-10}\exp(-82.7/T)$\\
B19$^+$ & $8.31\times 10^{-11}\exp(0.92/T) $\\
B19$^-$ & $1.02\times 10^{-9}\exp(-256/T) $\\
B20$^+$ & $1.68\times 10^{-10}\exp(0.77/T) $\\
B20$^-$ & $6.04\times 10^{-10}\exp(-88.8/T) $\\\tableline
\tablenotetext{a}{\cite{oka04}, \cite{hugo09}, \cite{coutens14} and references therein.}
\tablenotetext{b}{The unit of temperature ($T$) is the Kelvin.}
\end{tabular}

\end{center}
\end{table}

\clearpage



\end{document}